\documentclass[onecolumn,11pt,draftclsnofoot,peerreview]{IEEEtran}

\usepackage{verbatim}  

\usepackage{subeqnarray} 
\usepackage{amsfonts}
\usepackage{amsmath}
\usepackage{amssymb}
\usepackage{dblfloatfix}  
\usepackage[latin1]{inputenc}
\usepackage[english]{babel}

\usepackage{graphicx}
\usepackage{subfigure} 

\usepackage{pict2e}  

\usepackage{epsfig}

\newcommand{\ls}[1]
   {\dimen0=\fontdimen6\the\font
    \lineskip=#1\dimen0
    \advance\lineskip.5\fontdimen5\the\font
    \advance\lineskip-\dimen0
    \lineskiplimit=.9\lineskip
    \baselineskip=\lineskip
    \advance\baselineskip\dimen0
    \normallineskip\lineskip
    \normallineskiplimit\lineskiplimit
    \normalbaselineskip\baselineskip
    \ignorespaces
   }

\newcommand{\nc}{\newcommand}

\nc {\yt} {y^0(t)}
\nc {\wpp} {\omega_p}
\nc {\vsc} {v_s^*}
\nc {\wz} {\omega_z}
\nc {\vsm} {\left. \vsc \right|_{min}}

\nc {\wrr} {\omega_r}      

\nc {\xt} {x^0(t)}

\nc {\ws} {\omega_s}

\nc {\wc} {\omega_c}
\nc {\w} {\omega}

\nc {\vr} {v_r}
\nc {\vs} {v_s}
\nc {\vo} {v_o}
        \nc {\vrn} {v_{rn}}

\nc {\vc} {v_c}

\usepackage[usenames,dvipsnames]{color}

\nc {\ma} {m_a}

\nc {\phpd} {m_a}

\nc {\dc} {D}
\nc {\vm} {V_m}

\nc {\gv} {G_{v}(s)}
\nc {\gi} {G_{i}(s)}
\nc {\gc} {G_{c}(s)}
\nc {\gp} {G_{p}(s)}

\nc {\myca} {\mathcal{C}_1}
\nc {\mycb}{\mathcal{C}_2}
\nc {\mycc}{\mathcal{C}_3}
\nc {\mycd}{\mathcal{C}_4}
\nc {\myce}{\mathcal{C}_5}
\nc {\mycf}{\mathcal{C}_6}
\nc {\mycg}{\mathcal{C}_7}
\nc {\mych}{\mathcal{C}_8}
\nc {\myci}{\mathcal{C}_9}
\nc {\mycj}{\mathcal{C}_{10}}
\nc {\al} {\alpha(D,p)}
\nc {\az} {\alpha_0(D)}
\nc {\ab} {\alpha_1(D)}
\nc {\ac} {\alpha_2(D)}

\nc {\azz} {\alpha(D,0)}
\nc {\co} {c(D,p)}

\newcommand{\csch}{\mathrm{csch}}

\nc {\zc} {\omega_z}

\nc {\myk} {\theta}
\nc {\myrho} {\rho}
\nc {\mrho} {\eta}

\nc {\myt} {\mathcal{T}} 
\nc {\mys} {\mathcal{S}} 
\nc {\myl} {\mathcal{L}} 

\nc {\va} {v_\mathrm{a}}
\nc {\vf} {v_\mathrm{f}}

\nc {\vq} {v_L}

\nc {\kmax} {K_\mathrm{max}}
\nc {\mykk} {\mathrm{K}} 
\nc {\mykmax} {\mathrm{K}_\mathrm{max}} 

\nc {\wesr} {\omega_r}
\nc  {\mv} {m_v}

\newcounter{example}
\newenvironment{example}[1][]{\refstepcounter{example}\par\medskip\noindent%
   \textit{Example~\theexample. #1} \rmfamily}{\medskip}

\nc {\ssa} {\mathbb{S}_1}
\nc {\ssb} {\mathbb{S}_2}
\nc {\ssc} {\mathbb{S}_3}
\nc {\ssd} {\mathbb{S}_4}

\usepackage[noadjust]{cite}

\newenvironment{mydescription}[1]
  {\begin{list}{}%
   {\renewcommand\makelabel[1]{##1 \hfill}%
   \settowidth\labelwidth{\makelabel{#1}}%
   \setlength\leftmargin{\labelwidth}
   \addtolength\leftmargin{\labelsep}}}
  {\end{list}}

\nc {\mytau} {\epsilon}
\nc {\mytaub} {\delta}
\nc {\myf} {\mathcal{F}}
\nc {\wq} {\omega_q}
\begin{document}

\title{Unified Subharmonic Oscillation Conditions for Peak or Average Current Mode Control}

\author{Chung-Chieh {Fang}
\thanks{arXiv:1310.7433v1: Oct. 28, 2013; arXiv:1310.7433v2: \today.}\\   
\thanks{The author is with 
Sunplus Technology, 
Hsinchu 300, Taiwan (email: fangcc3@yahoo.com).}
}

\date{}
\maketitle

\begin{abstract}

This paper is an extension of the author's recent research in which only buck converters were analyzed. Similar analysis can be equally applied to other types of converters. In this paper, a unified model is proposed for buck, boost, and buck-boost converters under peak or average current mode control to predict the occurrence of subharmonic oscillation. Based on the unified model, the associated stability conditions are derived in closed forms. The same stability condition can be applied to buck, boost, and buck-boost converters. Based on the closed-form conditions, the effects of various converter parameters including the compensator poles and zeros on the stability can be clearly seen, and these parameters can be consolidated into a few ones. High-order compensators such as type-II and PI compensators are considered. Some new plots are also proposed for design purpose to avoid the instability. The instability is found to be associated with large crossover frequency. A conservative stability condition, agreed with the past research, is derived. The effect of the voltage loop ripple on the instability is also analyzed.

\end{abstract}

\newpage \tableofcontents \newpage


\section{Nomenclature}
\begin{mydescription}{long llllllllllllllllllllll}  
\item[{\color{Red}{$D$}}] duty cycle (Note: dimensionless parameters are highlighted in {\color{Red}{red}})
\item[$T$] switching period 
\item[$f_s=1/T$] switching frequency (unit: Hertz)
\item[$\ws=2 \pi f_s$] angular switching frequency (unit: rad/s)
\item[$\wc$] crossover frequency
\item[$\vs$] source (input) voltage
\item[$\vo$] output voltage
\item[$\vr$] reference voltage
\item[$\vc$] control voltage (at the output of the voltage-loop compensator)
\item[$v_d$] voltage across the diode
\item[$i_L$] 
inductor current, a triangular wave in the time domain 
\item[$v_L=L (d i_L/dt)$] 
voltage across the inductor, a square wave in the time domain
\item[$m_1$] inductor current slope $d i_L/dt$ when the switch is on
\item[-$m_2$] (negative) inductor current slope $d i_L/dt$ when the switch is off
\item[$\va=v_h-v_l$] amplitude of 
$v_L$ (also note: $\va=L(m_1+m_2)$)
\item[$v_h=L m_1$] the high value of $v_L$ 
\item[$v_l=-L m_2$] the low value of $v_L$ 
\item[$y(t)$] feedback signal for switching (the switch is turned off when $y \le h$, for example) %
\item[$h(t)$] PWM or stabilizing ramp signal
\item[$V_h$] the high value of the ramp 
\item[$V_l$] the low value of the ramp 
\item[$\vm=V_h-V_l$] ramp amplitude
\item[$\ma=\vm/T=\vm f_s$] ramp slope
\item[$m_v$] the effect on the required ramp slope $\ma$ due to the (added) voltage loop ripple 
%
\item[$L$] inductance
\item[$C$] capacitance
\item[$C_3$] capacitance of a (ceramic) capicitor (with small ESR) in parallel with $C$
\item[$R$] load resistance
\item[$R_s$] sensing resistance (for the inductor current $i_L$)
\item[$R_c$] equivalent series resistance (ESR) of $C$
\item[{\color{Red}{$\rho= R/(R+R_c)$}}] a dimensionless parameter to show the effect of $R_c$ ($\rho=1$ if $R_c=0$)
\item[$\mytau = {\color{Red}{\mytaub}} T$] switching delay
\item[{\color{Red}{$\myt(s)=\myt_i(s)+\myt_v(s)$}}] loop gain (with two parts contributed by the current and voltage loops)
\item[{\color{Red}{$\myt'(s)=e^{-s \mytau} \myt(s)$}}] loop gain with a switching delay $\mytau := \mytaub T$
\item[{\color{Red}{$\myt_i(s)$}}] the part of loop gain contributed by the current loop
\item[{\color{Red}{$\myt_v(s)$}}] the part of loop gain contributed by the feedback voltage loop
\item[$\myca$, $\mycb$, $\mycc$, etc.] a case for a typical loop gain (see Table~\ref{13tab} or [4] 
for the case number) 
\item[$K_c$] compensator gain
\item[$\wpp$] compensator pole
\item[$\wz$] compensator zero
\item[$\wrr=1/R_c C$] ESR zero
\item[$\wq=\wesr(1+C/C_3)$] a pole contributed by adding $C_3$ in parallel with $C$ 
\item[{\color{Red}{$p=\wpp/ \ws$}}] normalized (by $\ws$) compensator pole
\item[{\color{Red}{$z=\wz/ \ws$}}] normalized compensator zero
\item[{\color{Red}{$r=\wrr/ \ws$}}] normalized ESR zero
\item[{\color{Red}{$q=\wq/ \ws$}}] normalized $\wq$
\item[{\color{Red}{$\gi=R_s /s  L$}}] current-loop integrator to convert $v_L$ to $R_s i_L$ (for modeling purpose only) 
\item[$\gc$] current-loop compensator transfer function
\item[$\gv$] voltage-loop compensator transfer function
\item[{\color{Red}{$K=\frac{\va R_s K_c}{\vm \zc L  \ws}$}}] gain of the current feedback loop for the type-II compensator case
\item[{\color{Red}{$\mykk=Kz$}}] gain of the current feedback loop for the PI compensator case 
\item[{\color{Red}{$\kmax$}}] maximum allowable $K$ to avoid subharmonic oscillation (i.e., need  $K < \kmax$)
\item[{\color{Red}{$\mykmax$}}] maximum allowable $\mykk$ to avoid subharmonic oscillation (i.e., need  $\mykk < \mykmax$)
\item[$K_v =\frac{\rho \vs K_c  }{ T L C \wz}$] gain of the voltage feedback loop for the type-II compensator case
\item[{\color{Red}{$k =\rho  K_c T/R_s  C \wz$}}] a dimensionless voltage feedback gain ($k=0$ if the voltage feedback loop is open)
\item[{\color{Red}{$\al$}}] 
a function used as a building block of most typical stability conditions 
\item[{\color{Red}{$\alpha_k(D)$}}] the $k$-th coefficient term of $\al=\sum\limits_{k=0}^\infty (-1)^k \alpha_k(D) p^k$ 
\item[{\color{Red}{$\co$}}] the  high order $k \ge 2$ (correction) terms of $\al$, i.e., $\co=\sum\limits_{k=2}^\infty (-1)^k \alpha_k(D) p^k$
\item[{\color{Red}{$k_p$}}] proportional feedback gain of the voltage loop
\item[{\color{Red}{$\myf(\myt(s) )$}}] an F-transform to convert a loop gain $\myt(s)$ to a stability condition, $\myf(\myt(s) ) < 1$
\item[$\mys =\myf (\phpd \myt(s) ) $] an S-plot to show the required stabilizing ramp {\em slope} (stability requires $\mys < m_a$)
\item[{\color{Red}{$\myl =\myf(\myt(s) )$}}] an L-plot which is an F-transform of a {\em loop} gain (stability requires  $\myl =\mys / m_a  < 1$) 
\item[$m_i =\myf (\phpd \myt_i(s))$] the part of the S-plot ($\mys=m_i+m_v$) contributed by the current loop
\item[$m_v =\myf (\phpd \myt_v(s))$] the part of the S-plot contributed by the voltage feedback loop
\end{mydescription}

\section{Introduction}

\begin{figure}[b]  
 \centering 
\includegraphics[153pt,359pt][434pt,461pt]{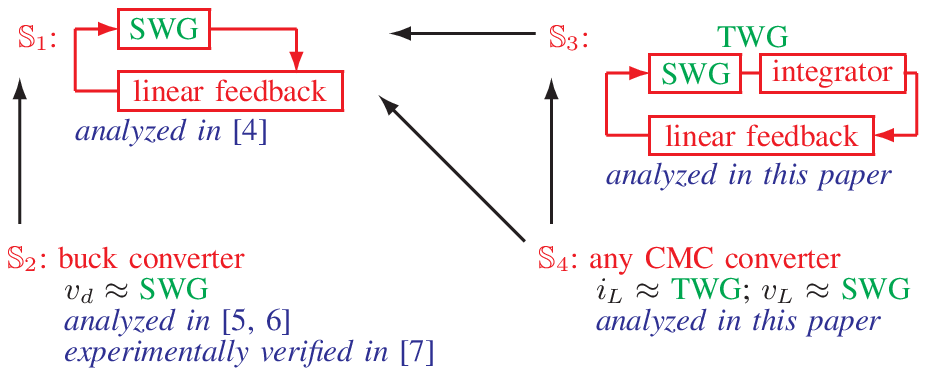}
\caption{The systems $\ssb$, $\ssc$ and $\ssd$ can be converted to $\ssa$ for further analysis.}
\label{ssa} 
\end{figure}

For DC-DC converters with current mode control 
(CMC) or voltage mode control (VMC), subharmonic oscillation (fast-scale instability, FSI) may occur \cite{jcmc,ctqls08}. 
The instability is common in 
peak CMC (PCMC), but rarely reported in average CMC 
(ACMC) \cite{slgz01}.  

Consider the following four closely related nonlinear systems: 
\begin{mydescription}{ll}
\item[$\ssa$:] a square wave generator (SWG) with a linear 
feedback; 
\item[$\ssb$:] a 
buck converter; 
\item[$\ssc$:] a triangular wave generator (TWG) with a linear 
feedback; and 
\item[$\ssd$:] any CMC converter. 
\end{mydescription}

The systems $\ssb$, $\ssc$ and $\ssd$ can be converted (denoted by ``$\to$'') to $\ssa$ 
as shown below (see also Fig.~\ref{ssa}):
\begin{mydescription}{long llllll}
\item[$\ssb \to \ssa$:] In the buck converter, the voltage 
$v_d$ across the diode (or the second switch) is a square 
wave, then $\ssb \to \ssa$ 
\cite{jhb12,jnewhb}. 
\item[$\ssc \to \ssa$:] A TWG is equivalent to an SWG plus an 
integrator, and an integrator plus a linear feedback is still a linear feedback, then $\ssc \to \ssa$. 
\item[$\ssd \to \ssa$:] In CMC, the inductor current 
$i_L$ is a triangular wave (like an output of TWG), then $\ssd \to \ssc \to \ssa$,
which makes a unified CMC model possible.  
\item[$\ssd \to \ssa$:]  Also, the voltage 
$v_L=L 
(d i_L /dt)
$ across the inductor is a square wave, then 
$\ssd \to \ssa$. 
\end{mydescription}

\begin{figure}[t] 
 \centering 
\includegraphics[163pt,381pt][424pt,441pt]{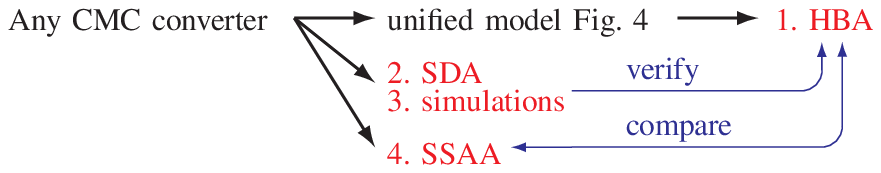}
\caption{Throughout the paper, the CMC converter is analyzed in four ways: it is first analyzed by HBA, independently verified by SDA and simulations, and  then compared with SSAA.}
\label{mythree} 
\end{figure}

Although harmonic balance analysis (HBA)  \cite{jhb12,jnewhb,jnewhb_cot} 
has been applied to buck converters ($\ssb$ 
or $\ssa$) to obtain 
the FSI conditions, and experimentally verified in  \cite{jdivider}, its
application to any CMC converter has not been reported. 
Based on the FSI conditions for $\ssa$, this paper derives the 
general FSI conditions for $\ssc$ 
and $\ssd$. As shown in Fig.~\ref{mythree}, 
all of the results are independently verified by time-domain simulations and sampled-data analysis (SDA) 
\cite{fa01}, a known accurate analysis for DC-DC converters. 
FSI occurs when a sampled-data (discrete-time) pole 
crosses -1. The results are also compared with state-space average analysis (SSAA) which is less accurate. 

This paper focuses on the FSI conditions and tries to 
answer the following questions: 
\begin{enumerate}
\item The buck and boost converters have different 
dynamics. For example, the boost converter has 
a right half plane zero \cite{em01}. Do these two converters with CMC 
essentially have the same dynamics? 
\item In the past research \cite{r91} on PCMC, a sampling effect is included 
in order to predict FSI, which requires 
increasing the system dimension. However, its 
application to ACMC has been questioned  \cite{sb99,ylm13}. 
Also, the ramp in PCMC is used for stabilization, 
whereas the ramp in ACMC is used for PWM 
modulation. Does a unified CMC model, applicable 
to both PCMC and ACMC, exist without 
increasing the system dimension? 
\item Is the unified CMC model also applicable to buck, 
boost, and buck-boost converters? 
\item A converter has many parameters. Each parameter 
has a different effect. Can these parameters be 
consolidated into a few parameters to predict FSI? 
Is there a {\em single} plot which predicts FSI?
\end{enumerate}
The answers to these questions will be shown to be 
affirmative. 

For PCMC with open voltage loop, the FSI conditions have been well reported. 
For ACMC 
\cite{slgz01,sb99,ylm13,d90,tlr93,slc99,c00,l08,lolb09,lolb11,ylm11b},
however, no accurate {\em general} closed-form FSI 
conditions have been reported. Also, the effects of the 
compensator poles and zeros on the stability have also 
not been reported. In this paper, the closed-form FSI 
conditions are derived, and the effects of many converter 
parameters can be clearly seen. 

The remainder of the paper is organized as follows. 
The FSI conditions based on harmonic balance analysis \cite{jhb12} are reviewed in Section~\ref{sec:hba}.
A unified CMC model is proposed in Section~\ref{sec:uni}. It 
is then applied to various PCMC and ACMC schemes in 
Section~\ref{sec:app}. The effect of the voltage loop ripple (at the output of the voltage-loop
compensator) is considered 
in Section~\ref{13sec:vloop}. Conclusions are collected in Section~\ref{13sec:conclu}.

\section{Review of FSI Conditions Based on 
Harmonic Balance Analysis} \label{sec:hba} 

FSI conditions based on harmonic balance analysis \cite{jhb12} are briefly reviewed.
Consider a unity-gain SWG with a linear feedback. 
Denote the switching period as $T$ and the switching frequency as $f_s=1/T$, and let $\ws=2 \pi f_s$.
Let the linear feedback transfer function be $\myt(s)$.
Let $\wpp$ be the pole and $\wz$ be the 
zero, for example. 
 Let $ p = \wpp/\ws$ and 
$ z = \wz/\ws$. Let $D$ be the duty cycle.
Take $\myt(s)= \ws/(s + \wpp)$, for example. 
The stability condition (to avoid FSI) is $\al < 1$, where
\begin{eqnarray*}
\al &=& 
2 \pi \csch(2 \pi p)- \pi e^{ \pi p (1-2D)} \csch( \pi p) 
\\
&:=& \sum\limits_{k=0}^\infty (-1)^k \alpha_k(D) p^k\\
&:=& \az - \ab p + \co \\
\az &=&\pi (2D-1) 
\\
\ab &=&  \pi^2 (2D^2-2D+1) 
\end{eqnarray*}
For other typical loop gains \cite{jhb12}, see Table~\ref{13tab}.
Based on partial fraction 
decomposition 
of $\myt(s)$, most FSI conditions are related 
with $\al$, which is a building block of other FSI 
conditions \cite{jhb12}. It is also the reason why a special form 
of $\al$ is defined as above. 

\begin{table}[t] 
\caption{Stability condition for typical loop gains $\myt(s)$ \cite{jhb12}.} 
\centering 
\begin{tabular}{cll}
	\hline
Case & $\myt(s)$ & Stability condition to avoid FSI\\
	\hline
$\myca$ & $\frac{1}{s+\wpp}$ & $\frac{1}{\ws}\al 
< 1
$\\
$\mycb$ & $\frac{1}{s}$ & $\frac{1}{\ws}\az < 1$\\
$\myce$ & $\frac{1}{s(1+s/\wpp)} 
$ 
& 
$\frac{1}{\ws}(\az-\al) < 1$
\\
$\mycg$ & $\frac{1+s/\wz}{s^2} 
$ 
& $\frac{1}{\ws^2}(\frac{1}{z} \az + \ab) < 1$\\
$\myci$ & $\frac{1+s/\wz}{s^2(1+s/\wpp)} $ & $
\frac{1}{\ws^2}(\ab+(\frac{1}{p}-\frac{1}{z})(\al-\az)) < 1
$\\
	\hline
\end{tabular}
 \label{13tab}
\end{table}

\section{Unified PCMC/ACMC  Model 
for Different Converters} \label{sec:uni} 
\begin{figure}[!t]  
 \centering 
\includegraphics[147pt,334pt][449pt,508pt]{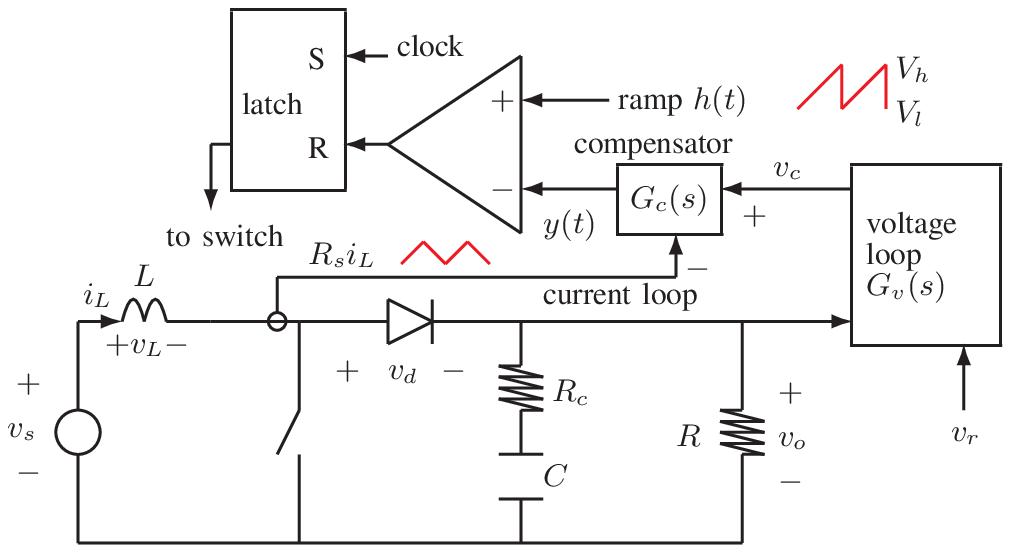}
\caption{A CMC boost converter with a current-loop compensator $\gc$ and a voltage-loop compensator $\gv$.}
\label{23sys_boo} 
\end{figure}

Consider a CMC boost converter shown in Fig.~\ref{23sys_boo}, where
%
$\gc$ is the current-loop compensator transfer function, $\gv$ is the voltage-loop compensator transfer function, 
$v_s$ is the source voltage,
$v_o$ is the output voltage, 
$\vc$ is the control voltage, 
$v_r$ is the reference voltage, 
$R_s$ is sensing resistance,
$y$ is 
a feedback signal, and
$h$ is a PWM or compensating 
ramp varying from $V_l$ to $V_h$.
In Sections~\ref{sec:uni} and \ref{sec:app}, $\vc$ is assumed constant.
The effect of voltage loop ripple is analyzed in Section~\ref{13sec:vloop}.
Let 
the equivalent series resistance (ESR) be $R_c$.
Denote the ramp slope as $m_a$ and the ramp amplitude as $\vm=V_h-V_l=m_a T$.

\begin{figure}[!t]  
 \centering 
\includegraphics[175pt,372pt][462pt,468pt]{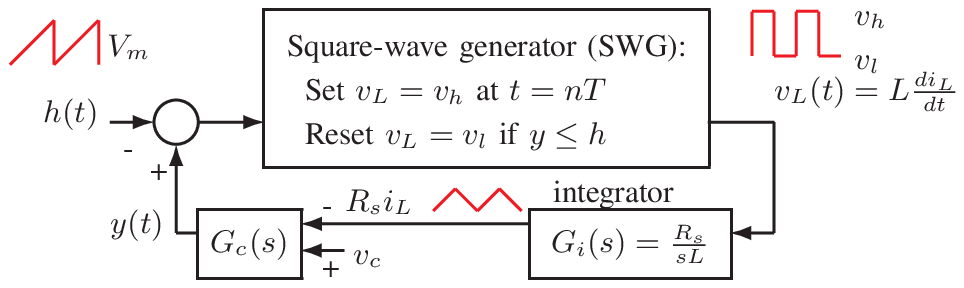}
\caption{
Unified CMC 
model, applicable to PCMC/ACMC buck, boost, and buck-boost converters, for example.}
\label{23unified} 
\end{figure}

\begin{figure}[!t]  
 \centering 
\includegraphics[168pt,656pt][435pt,731pt]{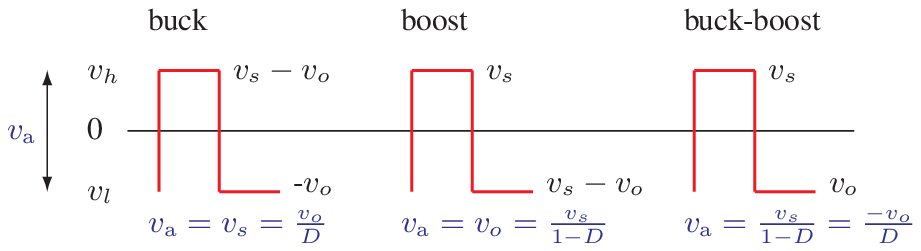}
\caption{The waveform of $v_L(t)=L \dot{i}_L$ for different converters.}
\label{23unified-n} 
\end{figure}

The inductor current $i_L$ is a triangular wave, and
and the 
voltage across the inductor
$\vq(t)=L d i_L/ dt$ is a 
square wave. 
Therefore, a CMC converter can be 
represented by a {\em unified} model shown in Fig.~\ref{23unified}.
Let the square wave $\vq(t)$ 
have a high value $v_h$, a low value $v_l$, and an amplitude $\va=v_h-v_l$, as shown in Fig.~\ref{23unified-n} for different converters.
Take the boost converter, for example.
When the switch is on, $L d i_L/ dt=\vs$.
When the switch is off, $L d i_L/ dt=\vs-\vo$.
Then, $\va=\vs-(\vs-\vo)=\vo=\vs/ (1-D)$. 
Denote 
$| \dot{i}_L|$ by $m_1$ 
and $m_2$ 
when the (first) switch is on and 
off, respectively. Then $v_h= L m_1$,
$v_l= -L m_2$,
$\va= L (m_1+m_2)$ which is another universal expression of $\va$.

\section{Applications to PCMC and ACMC} \label{sec:app} 
In Fig.~\ref{23unified}, the SWG contributes 
a gain $\va/ \vm$, then the (current) loop gain $\myt(s)= \va \gc \gi/ \vm$.
Different CMC schemes have different $\gc$, $\myt(s)$ 
and stability conditions (summarized in Table~\ref{23tab2}), which will be  verified by time-domain simulations (summarized in Table~\ref{22tab2}). 

\subsection{PCMC: Case $\mycb$}
In PCMC, $y = \vc -R_s i_L$, $\gc =1$ and $\myt (s)= \va \gi /\vm = \va R_s /s  \vm L$ which belongs to case $\mycb$. 
Let $\mys$ be an S-plot \cite{jhb12} to show the required stabilizing ramp slope. 
For $\mys < m_a$, the converter is stable.
From Table~\ref{13tab}, the stability condition is 
\begin{equation} \label{13eq:bc_cmc}
\frac{\va R_s \az }{\vm L \ws} < 1
\hspace{3mm} \mbox{or} \hspace{3mm}
\mys:=\frac{\va R_s (D-0.5)}{L} < \phpd
\end{equation}
where $\va 
= \vs$ for buck converters and $\va 
= \vs/(1-D)$ for boost or buck-boost converters, agreed with \cite{jasym}. 

\begin{table}[t] 
\caption{
Unified 
stability 
conditions (in terms of 
ramp slope $\phpd$) for different
CMC 
schemes, applicable to any CMC converter.} 
\centering 
\begin{tabular}{l}
	\hline 
\vspace{2mm}
{\bf PCMC} 
\hspace{2.2cm}
 $\mys:=\frac{\va R_s}{L}(\dc -\frac{1}{2})<\phpd$ \\ \vspace{2mm}
{\bf ACMC (type-II)} \hspace{.9cm}  
$\mys:=\frac{\va R_s K_c}{T  L  \ws^2}((\ab+(\frac{1}{p}-\frac{1}{z})(\al-\az))) <\phpd$\\
\hspace{.8cm} (for $\wz \ll \ws$) \hspace{.35cm}
$\mys:=\frac{\va R_s K_c}{T \zc L  \ws}(\az-\al) <\phpd$\\
\hspace{.8cm} (for $\wz \ll \ws$) \hspace{.35cm}
$ K := \frac{ \va R_s K_c}{\vm \zc L  \ws}
< \kmax(D,p) := \frac{1}{ \az-\al}$ \\
{\bf ACMC (PI)} 
\hspace{1.5cm} 
$\mys:=\frac{\va R_s K_c}{L}(\frac{2D-1}{2 \omega_z}
+ \frac{(1-2D+2D^2) T}{4} )
< \phpd  $ \\
\hspace{3.2cm}
$ \mykk := \frac{ \va R_s K_c}{\vm  L  \ws^2} < \mykmax(D,z) :=\frac{1}{ \az /z+\ab}$\\
{\em Note}: 
$\va 
= \vs$ for buck converters and $\va 
= \vs/(1-D)$ for boost or buck-boost converters.\\
\hline
\end{tabular}
 \label{23tab2}
\end{table}

\begin{table}[!t] 
\caption{
Stable/unstable boost converters in Examples 1-3.}
\centering 
\begin{tabular}{llllllll}
	\hline
Ex. & $K$ or $\mykk$ 
& D& $p$ or $z$ 
&Stability & In parameter space & Simulation & PM \\
\hline
\hline
1 & $K=0.4$ & {\color{Blue}{0.86}} & $p=0.75$ & {\color{Red}{unstable}} & [a] in Figs. \ref{c5sta}(d) \& \ref{c5pm}(d) & Fig. 8& {\color{Red}{$60^\circ$}} \\
 & $K=0.4$ & {\color{Blue}{0.85}} & $p=0.75$ & stable & [b] in Fig. \ref{c5sta}(d)& Fig. 10 &  \\
	\hline
2 & $K=1.3$ & 0.36 & {\color{Blue}{$p=0.17$}} & stable & [c] in Fig. \ref{c5sta}(c)& Fig. 11 &  \\
 & $K=1.3$ & 0.36 & {\color{Blue}{$p=0.18$}} & {\color{Red}{unstable}} & [d] in Figs. \ref{c5sta}(c) \& \ref{c5pm}(c) & Fig. 12 & {\color{Red}{$18^\circ$}} \\
 & $K=1.3$ & 0.36 & {\color{Blue}{$p=0.515$}} & {\color{Red}{unstable}} & [e] in Figs. \ref{c5sta}(c) \& \ref{c5pm}(c) & Fig. 14 & {\color{Red}{$33^\circ$}} \\
 & $K=1.3$ & 0.36 & {\color{Blue}{$p=0.52$}} & stable & [f] in Fig. \ref{c5sta}(c)& Fig. 16 &  \\
	\hline
3 & $\mykk=0.0232$ & {\color{Blue}{0.6}} & $z=0.018$ & {\color{Red}{unstable}} & [g] in Figs. \ref{c7sta}(c) \& \ref{Kmax_booPI} & Fig. 20& {\color{Red}{$89^\circ$}} \\
 & $\mykk=0.0232$ & {\color{Blue}{0.58}} & $z=0.018$ & stable & [h] in Figs. \ref{c7sta}(c) \& \ref{Kmax_booPI} & Fig. 22&  \\
	\hline
\end{tabular}
 \label{22tab2}
\end{table}

\subsection{ACMC with Type-II Compensator: Case $\myce$ or $\myci$} \label{sec:acmc2}
For ACMC, $y =\gc(\vc - R_s i_L)+ \vc$, which has 
an additional offset $\vc$ but does not affect the loop gain. 
Let the type-II phase-lead compensator (with $\wz <\wpp$) 
be 
\begin{equation} \label{eq:gc2}
\gc=\frac{K_c (1+s/ \wz)}{s (1+ s/ \wpp)}
\end{equation}
where $K_c$ is a gain.
Generally, $\wz \ll \ws$.
Let $K=\va R_s K_c/\vm \zc L  \ws$. 
Then
\begin{eqnarray}
\myt(s) &=&  
\frac{\va \gc \gi}{\vm}=
\frac{\va R_s K_c   (1+\frac{s}{ \wz})}{\vm L s^2 (1+ \frac{s}{ \wpp})  } \label{eq:highe}
\\
&\approx&  \frac{K \ws}{s(1+ \frac{s}{ \wpp})} \hspace{5mm} 
  \mbox{     (at frequency $\ws \gg \wz$)} \label{eq:high2}
\end{eqnarray}

\subsubsection{Based on SSAA: Converter is expected to be stable}
Let $\wc$ be the crossover frequency.
 Setting $|\myt(j \wc)|=1$ in (\ref{eq:high2}) leads to
\begin{eqnarray}
\wc &=&  
\frac{ \ws}{\sqrt{2}}  \sqrt{  \sqrt{   p^4+ 4K^2 p^2}   -p^2  } 
\\
&\approx & 
\left\{
\begin{array}{l}
\ws K 
\hspace{5mm} 
  \mbox{     (for    $K  \ll p$)} \label{eq:wcappro} 
\\
\\
\ws \sqrt{K p}  
\hspace{5mm} 
\mbox{      (for   $K  \gg p$)  } 
\end{array} \right.
\end{eqnarray}
A large $K$ leads to a large $\wc$.
From (\ref{eq:high2}), the phase margin (PM) is $90^{\circ} - \arctan(\wc/ \wpp) >0$. 
PM is a function of 
$K$ and $p$, independent of $D$.
For $K=100$, 2, 1.3, and 0.4, the plots of PM 
in the $(D,p)$ space are shown in Fig.~\ref{c5pm}, and the converter is expected to be always stable. 
As $K$ decreases, $\wc$ decreases and PM 
increases.

\subsubsection{Based on HBA: FSI may occur even with PM $>0$}
\begin{figure}[t]
    \centering 
\includegraphics[width=0.55\columnwidth]{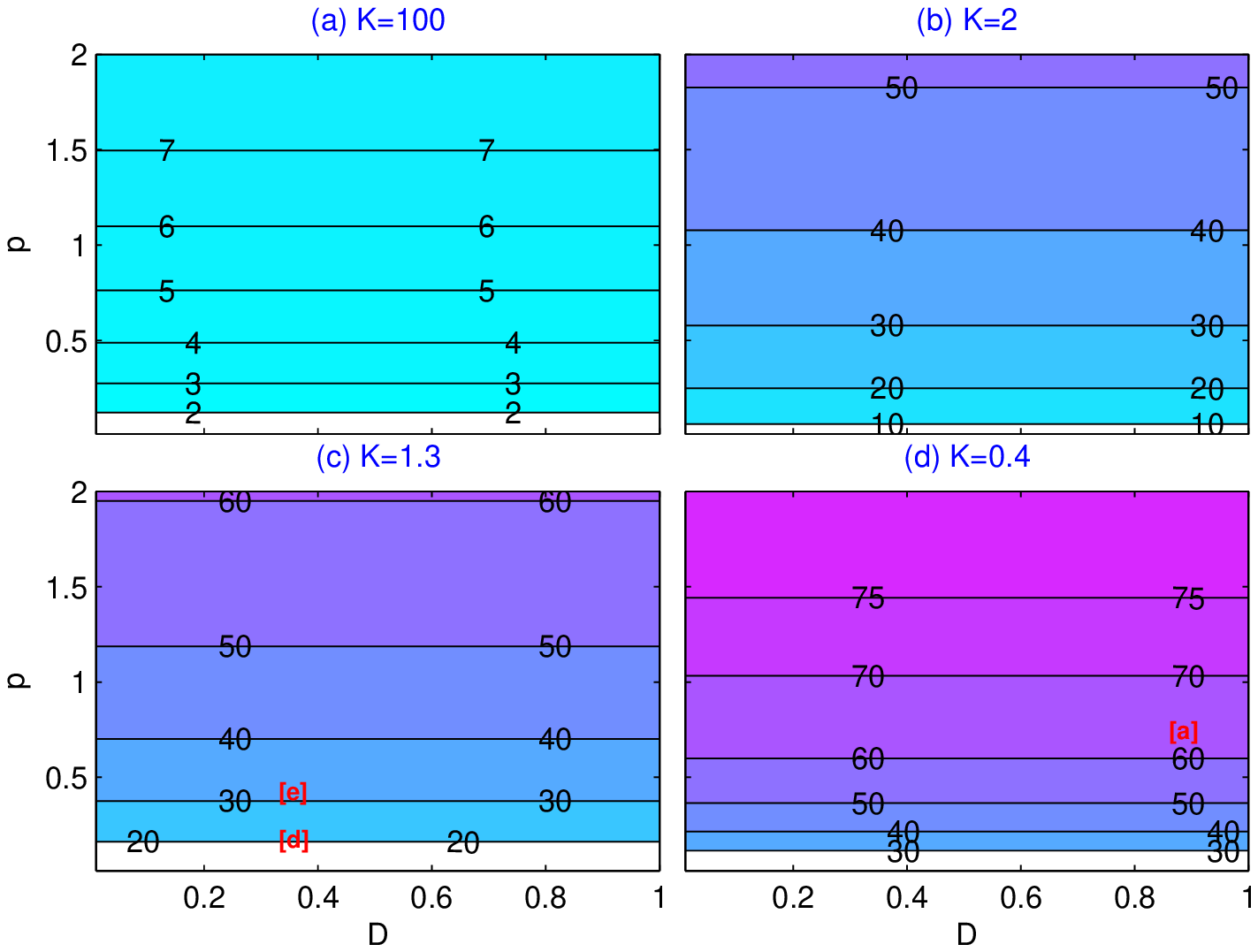}  
\caption{As $K$ decreases, PM
increases, independent of $D$, but FSI 
 still occurs as shown in Fig.~\ref{c5sta}, different colors for different PM.}
 \label{c5pm}
\end{figure}

\begin{figure}[t]
    \centering 
\includegraphics[width=0.55\columnwidth]{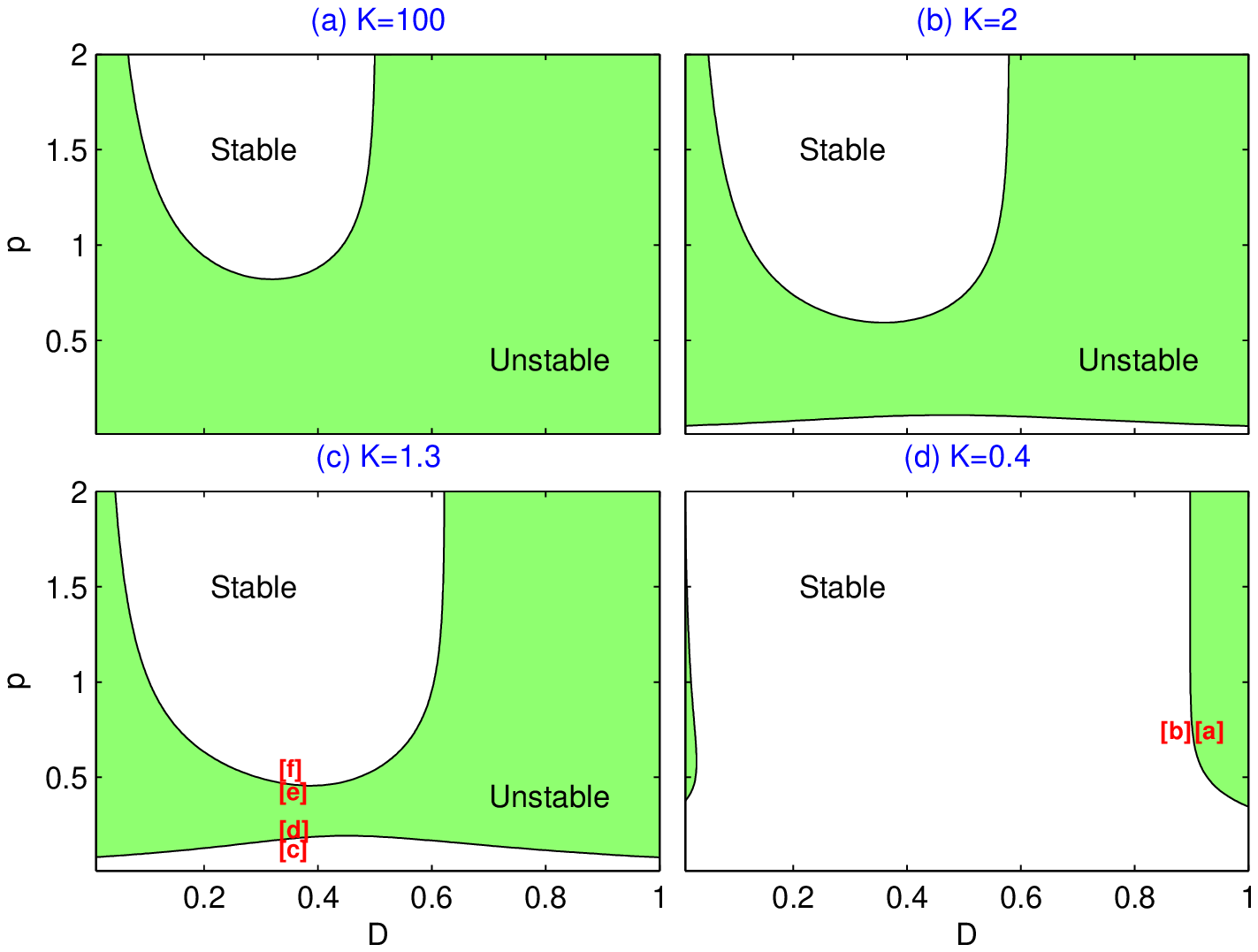}  
\caption{As $K$ decreases, the instability region still exists but shrinks.}
 \label{c5sta}
\end{figure}

From (\ref{eq:high2}), $\myt(s)$ belongs to case $\myce$ in  
Table~\ref{13tab}, and the stability condition to avoid FSI is 
\begin{equation} \label{13eq:bc_c5}
K( \az-\al) 
< 1
\end{equation}
which can be expressed in terms of the required ramp slope 
$\phpd$, as shown in Table~\ref{23tab2}.
FSI 
may occur if (\ref{13eq:bc_c5}) is not met.
For the same $K$ as in Fig.~\ref{c5pm}, the stable regions 
according to 
(\ref{13eq:bc_c5})
are shown in Fig.~\ref{c5sta}.
As $K$ decreases, the stability region enlarges, but there still exist instability regions.
From \cite{jnewhb}, no subharmonic oscillation occurs if $K < 1/ \pi$,
which is a conservative condition and it is approximately equivalent to $\wc < \ws/ \pi$ according to (\ref{eq:wcappro}).
A large $\wc$ leads to FSI.
However, such a condition $\wc < \ws/ \pi$ may be too conservative.
The converter can be designed according to the limit (\ref{13eq:bc_c5}) with 
larger $\wc$ for higher performance without losing stability.

Note that PM in Fig.~\ref{c5pm} is independent of $D$, whereas
the stability in Fig.~\ref{c5sta} depends on $D$.
Comparing Fig.~\ref{c5sta} with Fig.~\ref{c5pm}, one sees that the converter may be unstable even with 
PM = $60^{\circ}$,
for example, if $K=0.4$, $D=0.86$, and $p=0.75$ as shown in the next example.

\begin{figure}[t]
    \centering 
\includegraphics[width=0.55\columnwidth]{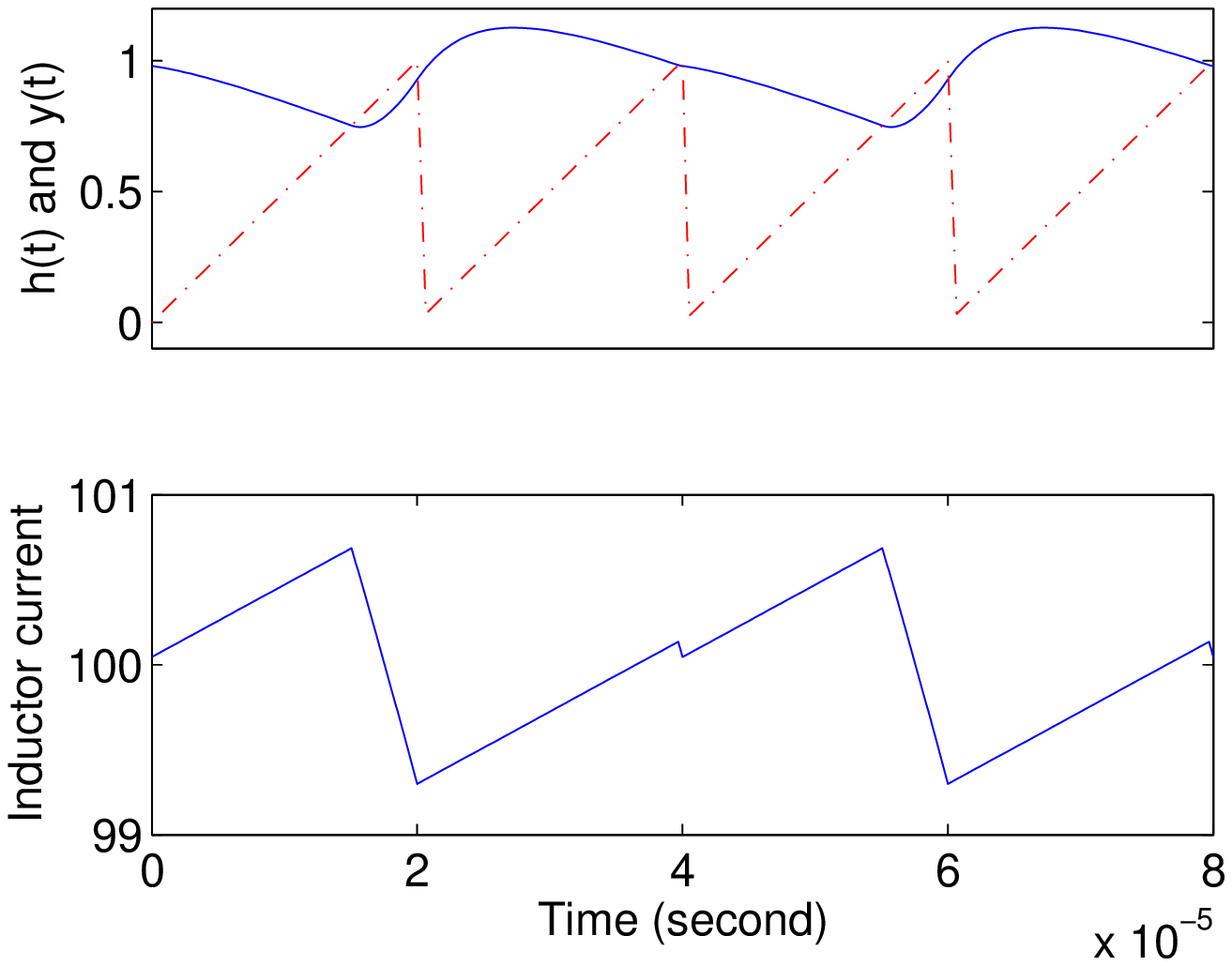}
\caption{The boost converter is unstable, $\vs=1.96$ V.}
 \label{boo_tp75}
\end{figure}

\begin{figure}[t]
    \centering 
\includegraphics[width=0.55\columnwidth]{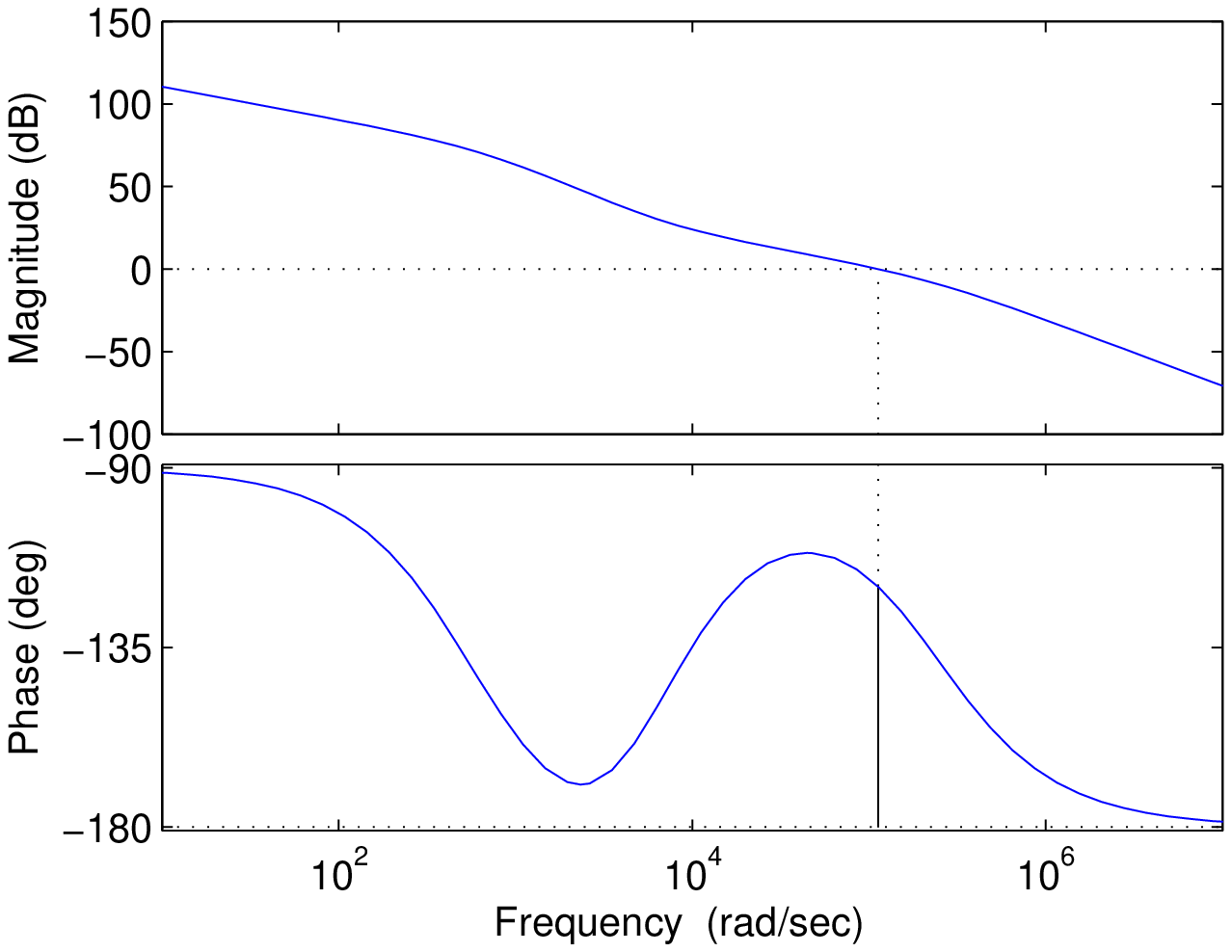}
\caption{The loop gain $\myt(j \w)$ has PM = $60^{\circ}$, $\vs=1.96$ V.}
 \label{boo_p75}
\end{figure}

\begin{figure}[t]
    \centering 
\includegraphics[width=0.55\columnwidth]{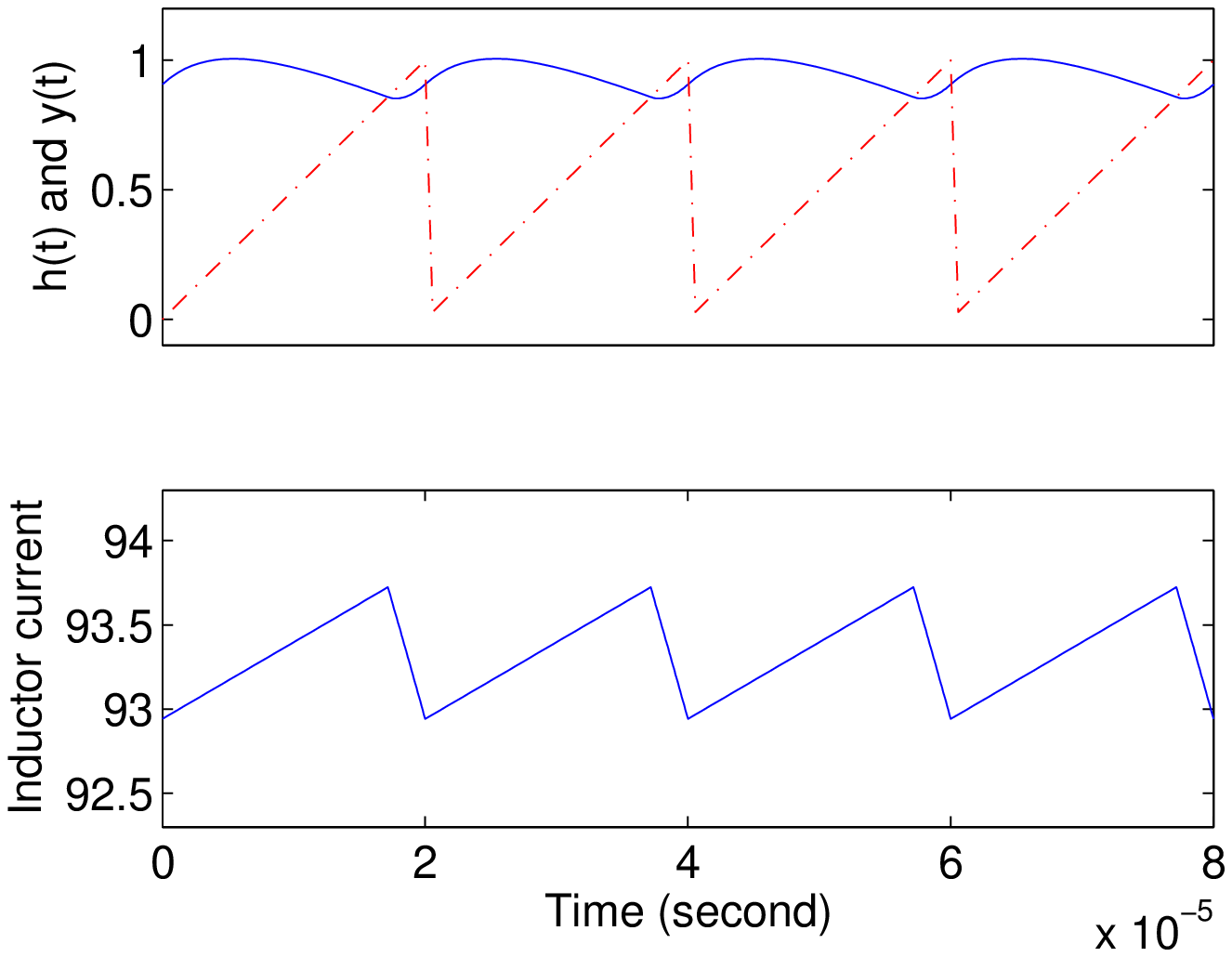}
\caption{The boost converter is stable, $\vs=2.1$ V.}
 \label{boo_tvs21}
\end{figure}

\vspace{2mm} \begin{example} 
({\em FSI with PM = $60^{\circ}$}.)
Consider an ACMC boost converter (adapted from the {\em buck} converter in \cite{jnewhb}) with a type-II compensator:
$v_o=14$ V,
 $\vm=1$ V,
$f_s=50$ kHz,
$L=46.1$ $\mu$H, $C=380$ $\mu$F, 
$R_c=0.02$ $\Omega$, 
$R_s=16.4$  m$\Omega$,
 $R=1$ $\Omega$,
$\zc=5652.9$ rad/s, $p=0.75$, $K_c=141670$, and $K=0.4$.

First,  let $\vs=1.96$ V and $\vc=1.64$ V. Here, $D=0.86$. The converter is unstable (Fig.~\ref{boo_tp75}) although its average model has PM = $60^{\circ}$ (Fig.~\ref{boo_p75}).
The linear 
average model is too simple to predict the FSI of the 
nonlinear converter.
Independent sampled-data analysis also shows 
an unstable pole at -1.02, and three stable poles at 0, 0.88, and 0.91, thus verifying the instability.

Next, let $\vs=2.1$ V and $\vc=1.53$ V. Now, $D=0.85$. The converter is stable (Fig.~\ref{boo_tvs21}).
In Fig.~\ref{c5sta}(d), for $K=0.4$, draw a line at $p=0.75$.
The instability occurs indeed around $D=0.86$.
\hfill  $\Box$  \end{example}

As reported in \cite{jnewhb}, the ACMC {\em buck} converter may have an unstable {\em window} of $p$.
The next example shows that the {\em boost} converter also has the same unstable window.

\begin{figure}[t]
    \centering 
\includegraphics[width=0.55\columnwidth]{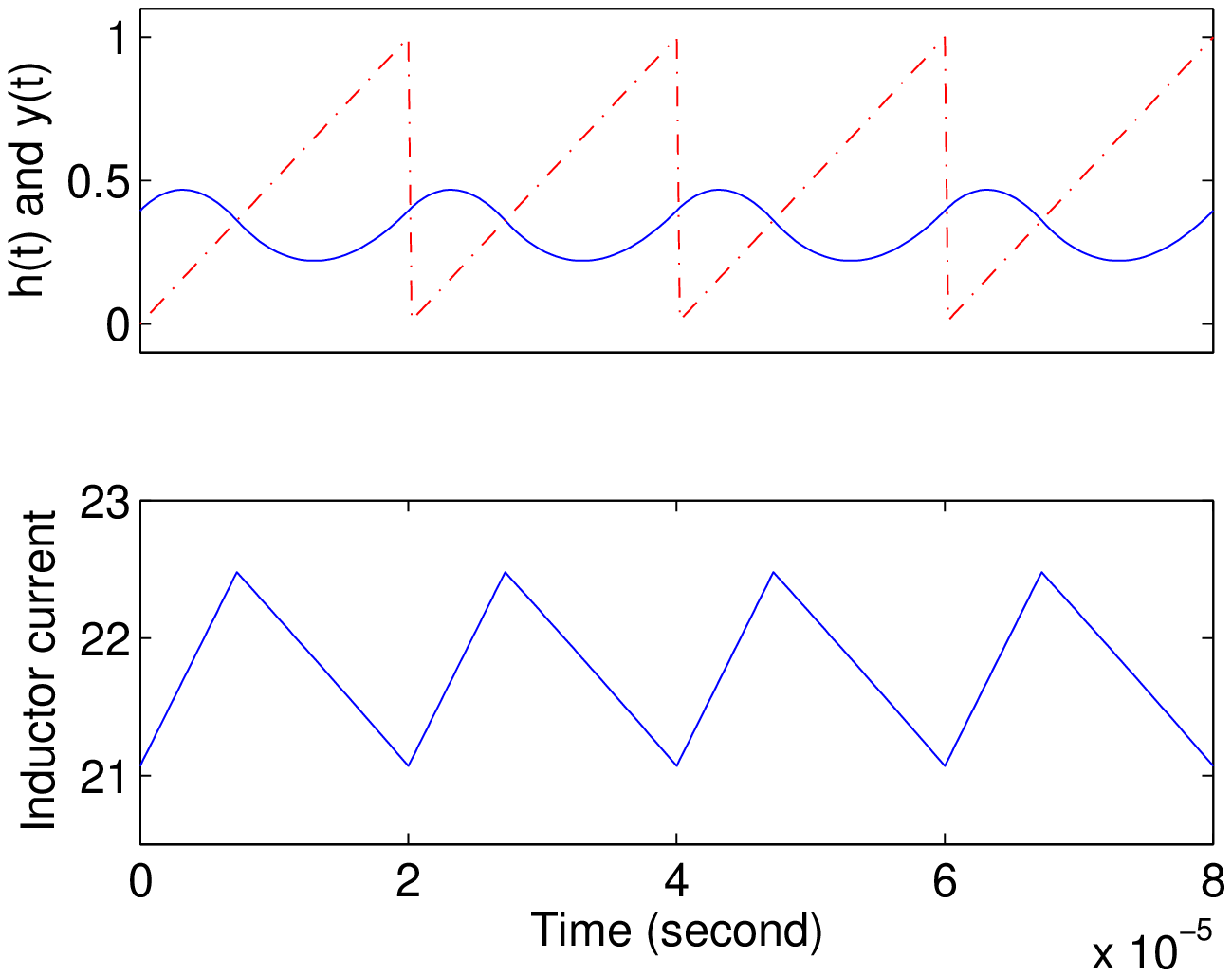}
\caption{The boost converter is stable, $p=0.17$.}
 \label{boo_tp17}
\end{figure}

\begin{figure}[t]
    \centering 
\includegraphics[width=0.55\columnwidth]{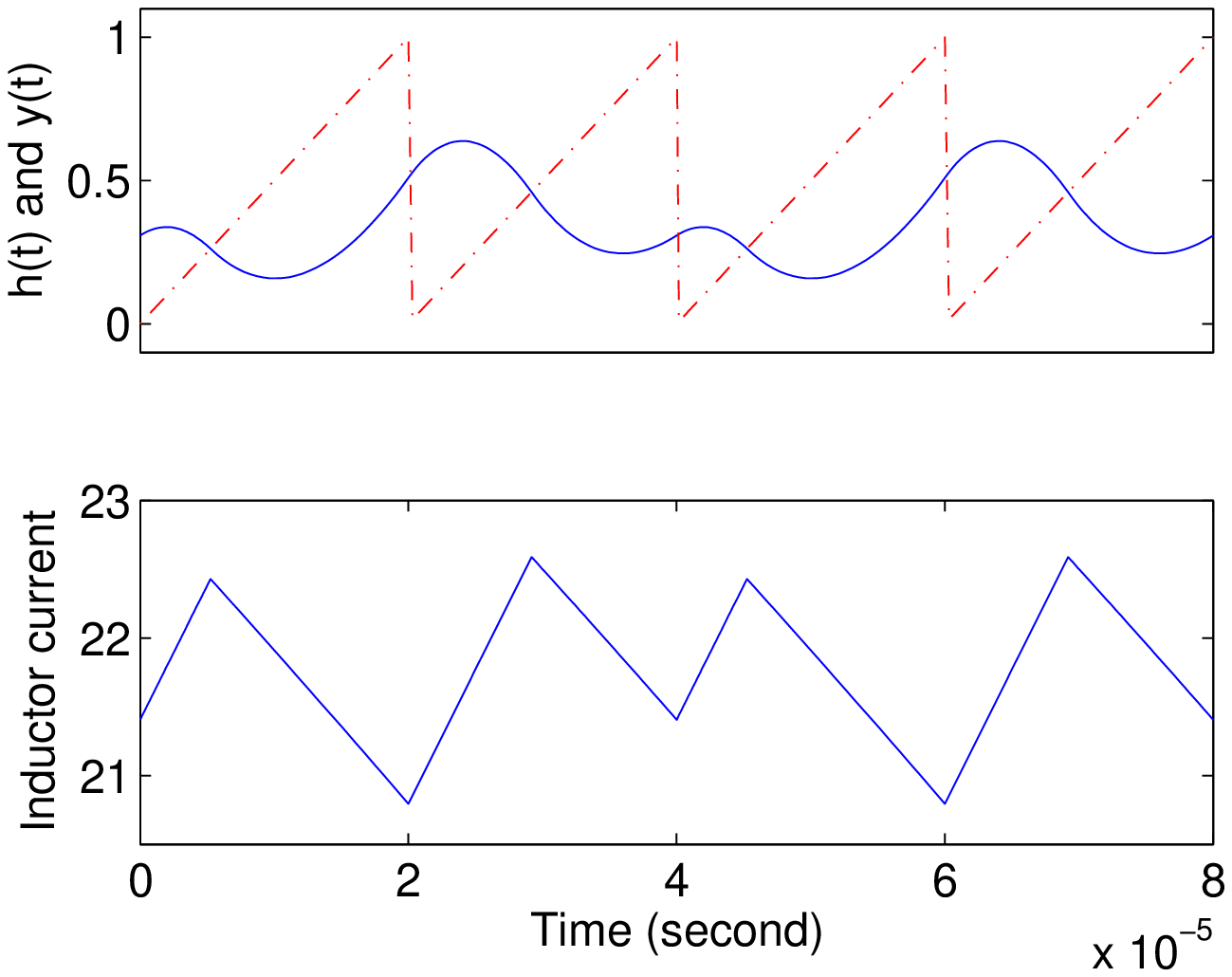}
\caption{The boost converter is unstable, $p=0.18$.}
 \label{boo_tp18}
\end{figure}

\begin{figure}[t]
    \centering 
\includegraphics[width=0.55\columnwidth]{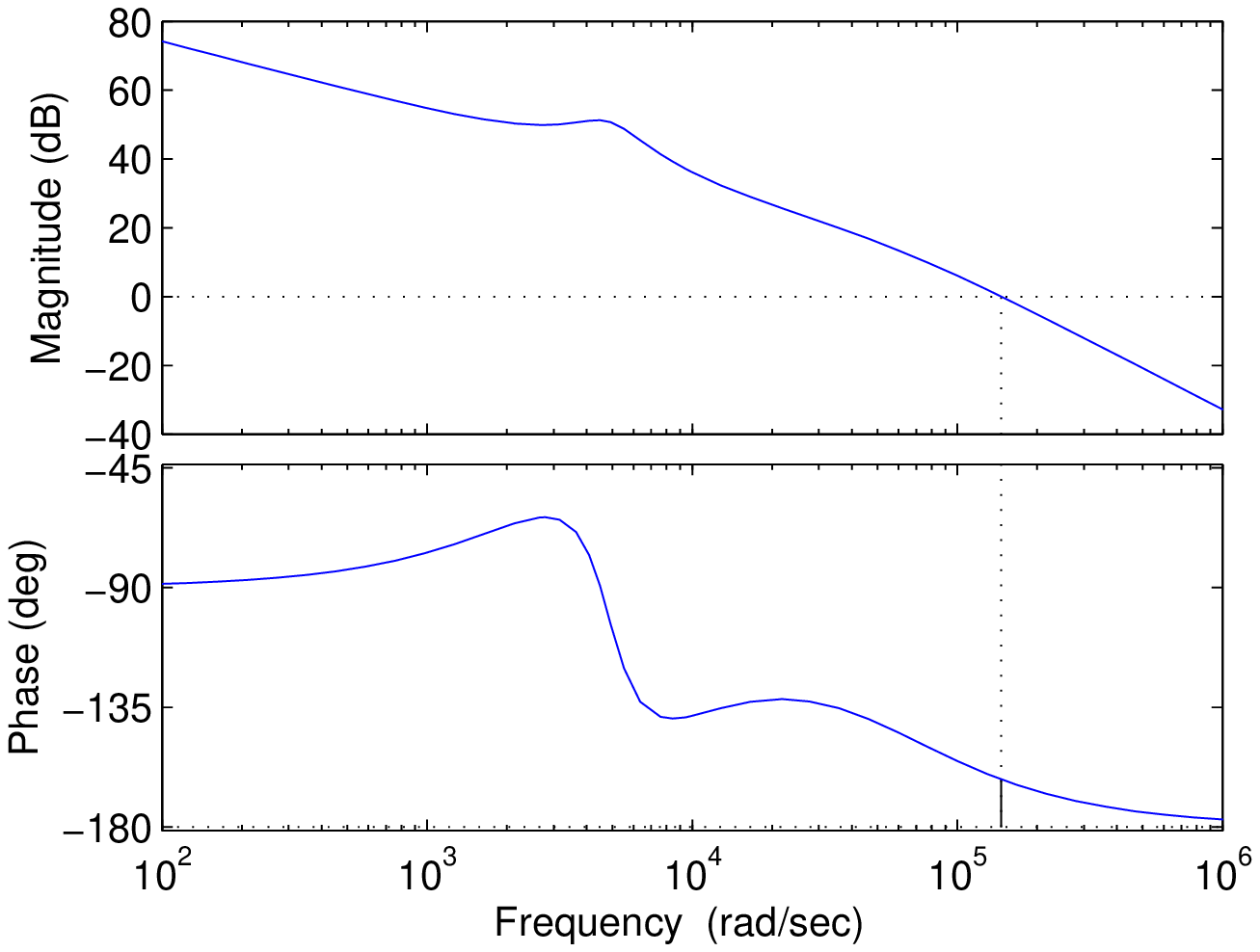}
\caption{The loop gain $\myt(j \w)$ has PM = $18^{\circ}$, $p=0.18$.}
 \label{boo_p18}
\end{figure}

\begin{figure}[t]
    \centering 
\includegraphics[width=0.55\columnwidth]{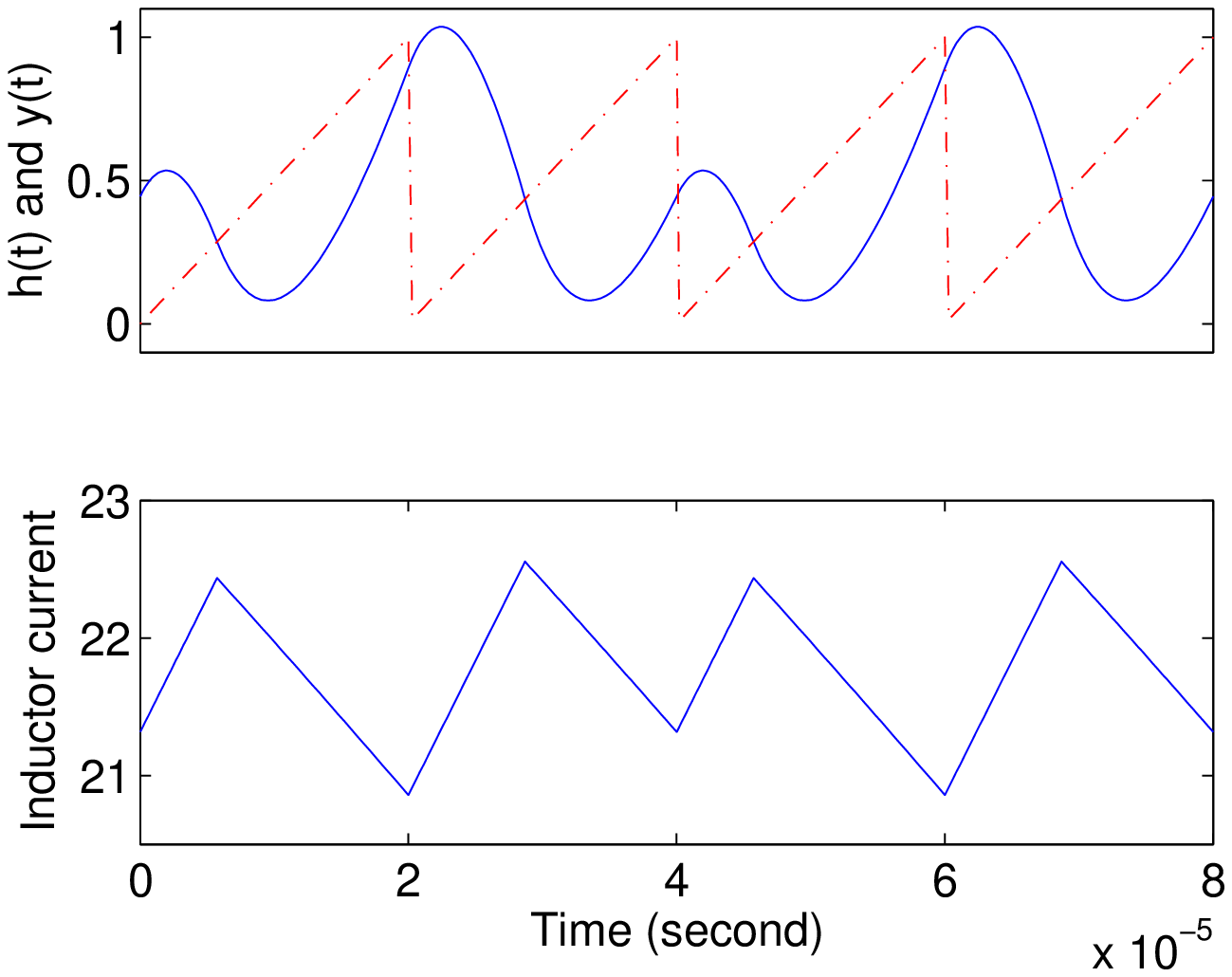}
\caption{The boost converter is unstable, $p=0.515$.}
 \label{boo_tp51}
\end{figure}

\begin{figure}[t]
    \centering 
\includegraphics[width=0.55\columnwidth]{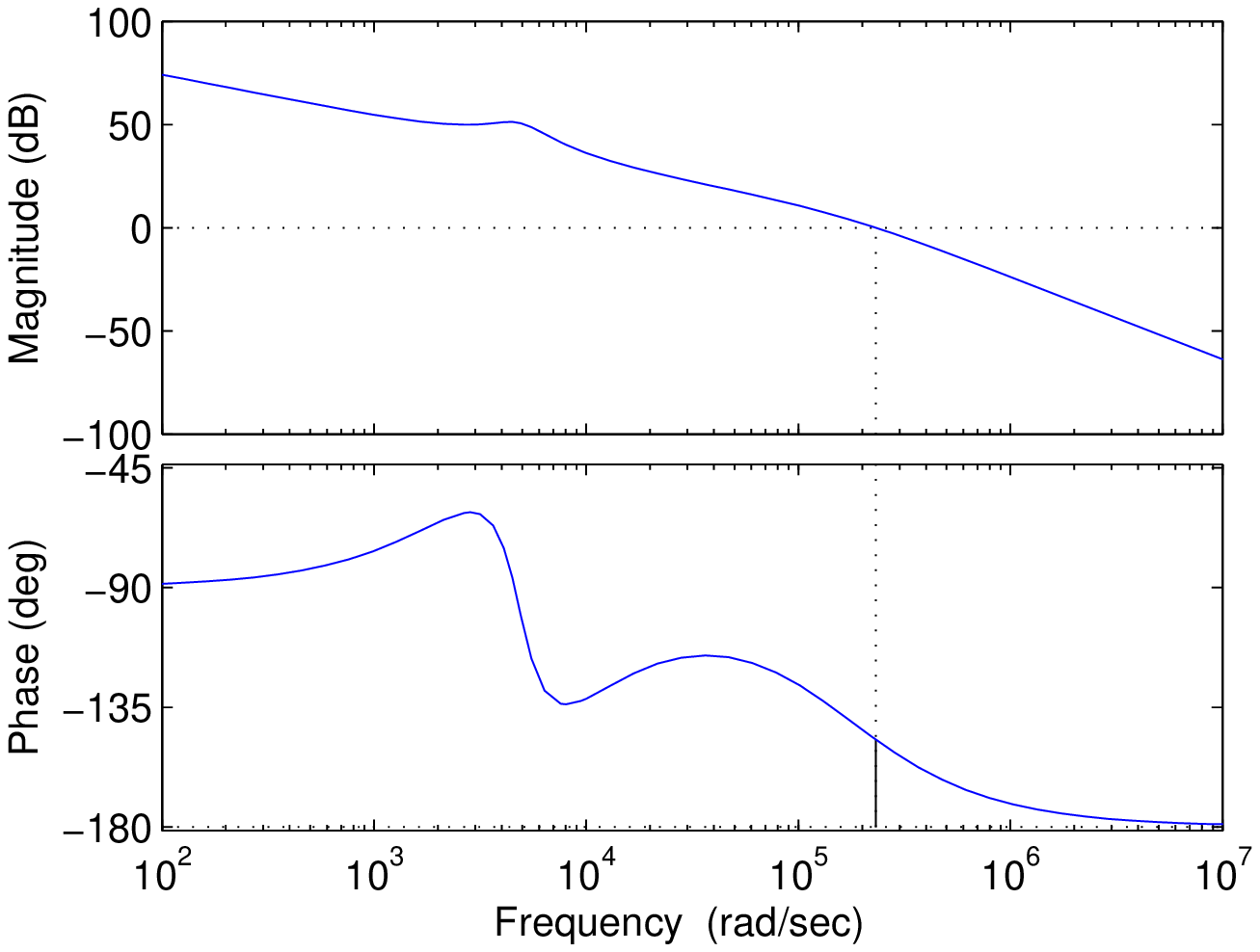}
\caption{The loop gain $\myt(j \w)$ has PM = $33^{\circ}$, $p=0.515$.}
 \label{boo_p51}
\end{figure}

\begin{figure}[t]
    \centering 
\includegraphics[width=0.55\columnwidth]{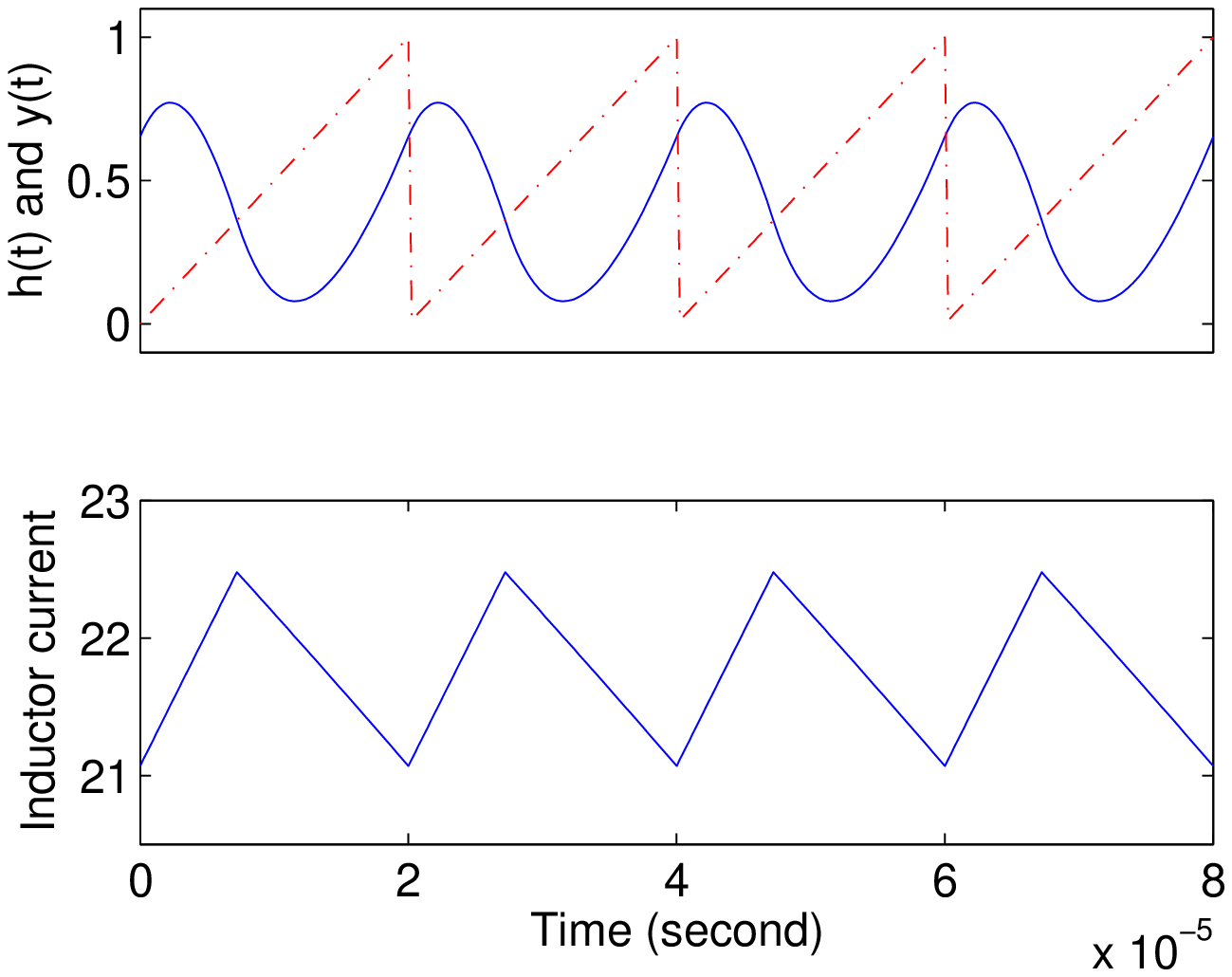}
\caption{The boost converter is stable, $p=0.52$.}
 \label{boo_tp52}
\end{figure}

\vspace{2mm} \begin{example} 
({\em Unstable window of $p$ in the boost converter, adapted from the buck converter in \cite{jnewhb}}.)
Continue from Example 1, but with
$\vs=9$ V,
$\vc=0.357$ V, 
and $K_c=460420$.
This {\em boost} converter example is actually adapted from Example 3 of \cite{jnewhb} for a {\em buck} converter, where an unstable window of $p$ was found.
This example 
also illustrates the buck and boost converters, known 
to have different dynamics, have the same occurrence 
of FSI with the same parameters. Both examples have 
the same $\va=14$ V, where $\va=\vs$ for the buck converter and $\va=\vo$ for the boost converter.
Both examples also have the same $D=0.36$ and $K=1.3$ (and also other parameters such as $R$, $L$, $C$, $R_c$, and $\wz$).
Therefore, an unstable window of $p$ for this boost converter is also expected. 
In Fig.~\ref{c5sta}(c), for $K=1.3$, draw a line at $D=0.36$, which 
shows an unstable window of $p \in [0.18, 0.515]$.
For $p < 0.18$ or $p > 0.515$, the converter is stable.

First, let $p=0.17$. The converter is stable (Fig.~\ref{boo_tp17}).

Second, let $p=0.18$. The converter is unstable (Fig.~\ref{boo_tp18}) although its average model has PM = $18^{\circ}$ (Fig.~\ref{boo_p18}).
Independent sampled-data analysis shows 
an unstable pole at -1.07, and three stable poles at 0.35, 0.88, and 0.91.

Third, let $p=0.515$. The converter is unstable (Fig.~\ref{boo_tp51}) although its average model has PM = $33^{\circ}$ (Fig.~\ref{boo_p51}).
Independent sampled-data analysis shows 
an unstable pole at -1.002, and three stable poles at -0.05, 0.88, and 0.91.

Fourth, let $p=0.52$. The converter is stable again (Fig.~\ref{boo_tp52}).
The boost converter indeed has an unstable window of $p \in [0.18, 0.515]$, same as the 
buck converter in \cite{jnewhb}.
\mbox{  } \hfill  $\Box$  \end{example}

Note that
$K=\va R_s K_c/\vm \zc L  \ws$, and one can see the effect of each parameter on the stability.
The condition (\ref{13eq:bc_c5}) can be expressed in terms of the required ramp slope $\phpd$, as shown in  Table~\ref{23tab2}:
\begin{equation} \label{13eq:bc_c5ma}
\phpd > \mys:=
 \frac{ \vo R_s K_c}{2\pi \zc L  } ( \az-\al) 
\end{equation}
The condition (\ref{13eq:bc_c5}) can be also expressed in terms of $K$:
\begin{equation} \label{13eq:bc_c52}
K 
< \frac{1}{ \az-\al}:=\kmax(D,p)
\end{equation}
if $\kmax(D,p)$ 
is positive. 
If $\kmax(D,p)$ is negative, the 
converter is always stable (because the inequality sign in (\ref{13eq:bc_c52}) is reversed and the condition (\ref{13eq:bc_c52}) is always
met).

\begin{figure}[t]
    \centering 
\includegraphics[width=0.55\columnwidth]{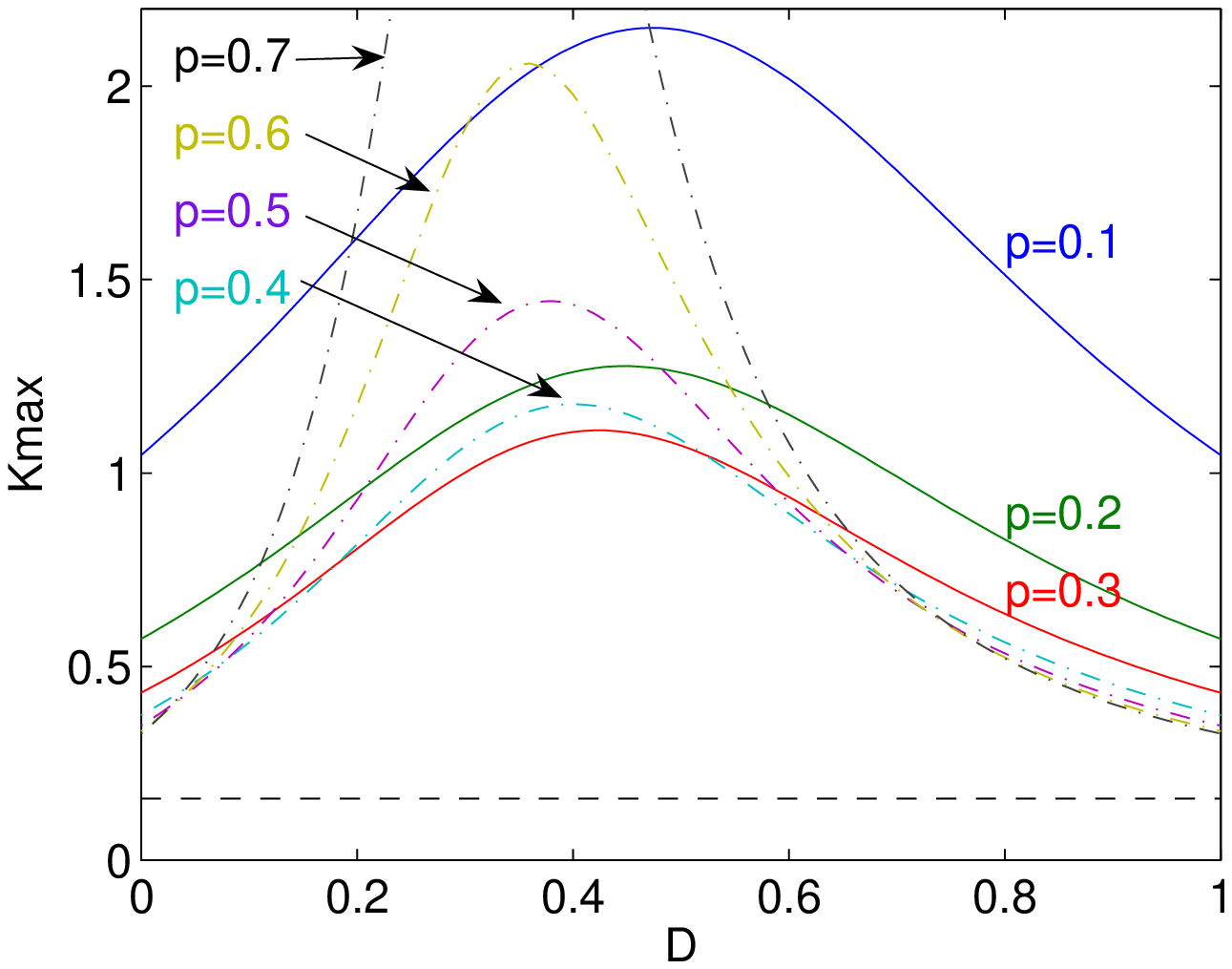} 
\caption{The plots of $\kmax(D,p)$ for the boost converter, dashed line for the conservative condition (\ref{13eq:sunboo}).}
 \label{Kmax_boo}
\end{figure}

For the {\em boost} converter, $\va=\vo$ which is fixed (if regulated), and
(\ref{13eq:bc_c52}) becomes
\begin{equation} \label{13eq:bc_c52boo}
K= \frac{ \vo R_s K_c}{\vm \zc L  \ws}
< \kmax(D,p)
\end{equation}
In \cite[Eq. 14]{sb99}, a conservative 
 condition was proposed:
\begin{equation} \label{13eq:sunboo}
\frac{ \vo R_s K_c}{\vm \zc L  \ws}
< \min [\frac{1}{\pi(1-D)}, \frac{1}{2 \pi}]=\frac{1}{2 \pi}
\end{equation}
where the effect of $p$ was neglected.
The plots of  (\ref{13eq:sunboo}) and $\kmax(D,p)$ for different values of $p$ are shown in Fig.~\ref{Kmax_boo}.

The plot of $\kmax(D,p)$ is quite nonlinear.
As $p$ increases from 0.1 to 0.3, $\kmax(D,p)$ decreases, whereas
as $p$ increases from 0.4 to 0.7 (and beyond), $\kmax(D,p)$ increases.
It indicates that around $p=0.3$ the stability region shrinks, agreed with Fig.~\ref{c5sta} which also indicates a possible unstable window of $p$.
One sees that the condition (\ref{13eq:sunboo}) reported in \cite{sb99} is conservative because $ \kmax(D,p) > 1/2 \pi $ as shown in Fig.~\ref{Kmax_boo}.

As indicated in Fig.~\ref{Kmax_boo}, the converter is prone to be stable around $D=0.4$ (than other values of $D$), 
also agreed with Fig.~\ref{c5sta}.
The closed-form $\kmax(D,p)$ is such nonlinear that it is difficult to further simplify it. 
Instead, one can make the plot of  $\kmax(D,p)$ to predict the stability. 

Since a {\em single} $K=\va R_s K_c/\vm \zc L  \ws$ contains {\em many} design parameters,
the plot of $\kmax(D,p)$ is very useful to design a stable converter.
Given the values of $p$ and the ranges of $D$,
one can make a plot of  $\kmax(D,p)$ and adjust different parameters so that the condition
 $K< \kmax(D,p)$ is met.

\begin{figure}[t]
    \centering 
\includegraphics[width=0.55\columnwidth]{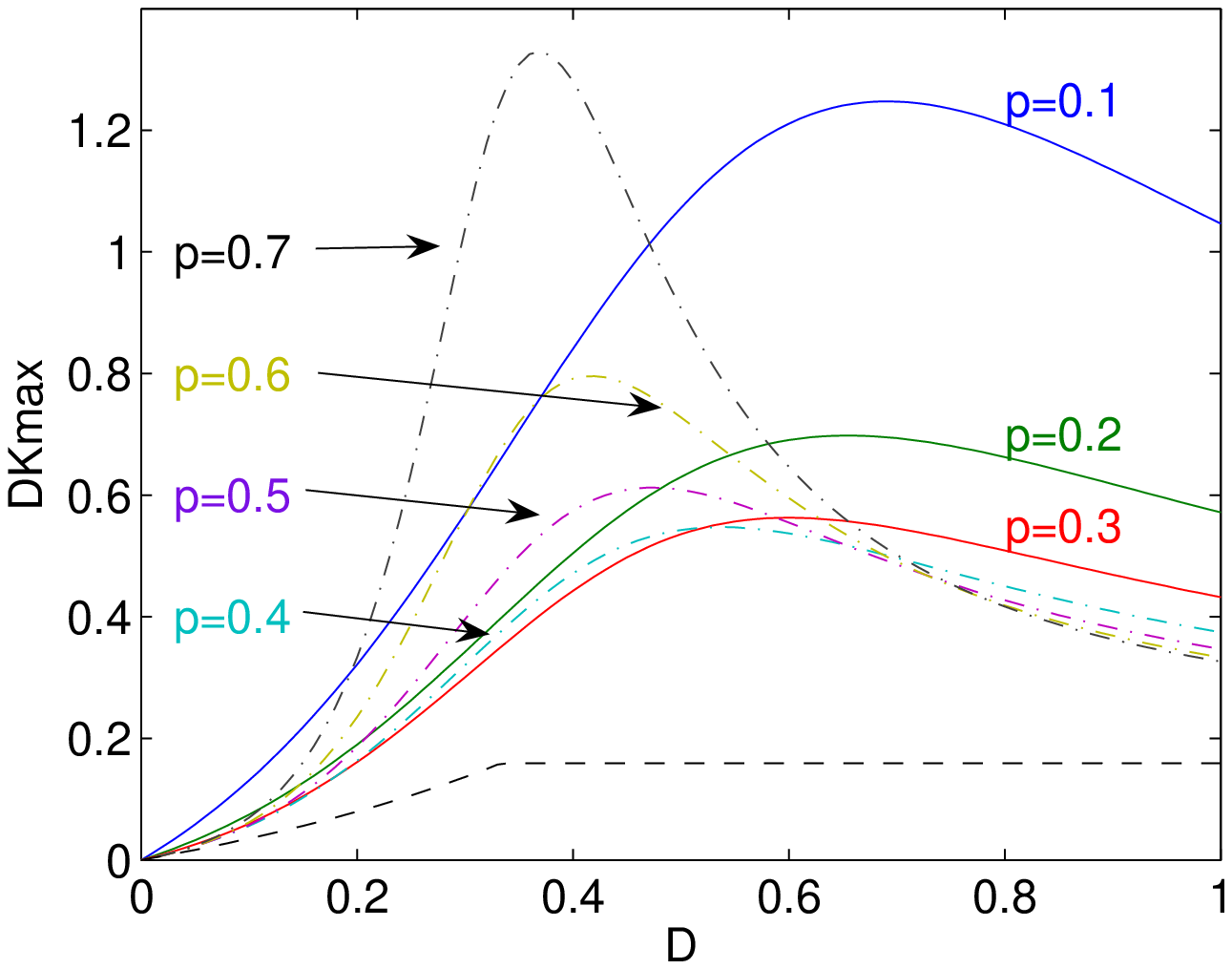} 
\caption{The plots of $D \kmax(D,p)$ for the buck converter, dashed line for the conservative condition (\ref{13eq:sunbuck}).}
 \label{Kmax_buck}
\end{figure}

For the {\em buck} converter, $\va=\vs=\vo/D$ and
(\ref{13eq:bc_c52}) becomes
\begin{equation} \label{13eq:bc_c52buck}
\frac{ \vo R_s K_c}{\vm \zc L  \ws}
< D \kmax(D,p)
\end{equation}
In \cite[Eq. 13]{sb99}, a conservative 
condition was proposed:
\begin{equation} \label{13eq:sunbuck}
\frac{ \vo R_s K_c}{\vm \zc L  \ws}
< \min [\frac{D}{\pi(1-D)}, \frac{1}{2 \pi}]
\end{equation}
The plots of  
(\ref{13eq:sunbuck}) and $D\kmax(D,p)$
for different values of $p$ are shown in Fig.~\ref{Kmax_buck}.
As $p$ increases from 0.1 to 0.3, $D\kmax(D,p)$ decreases.
As $p$ increases from 0.4 to 0.7, $D\kmax(D,p)$ increases.
One sees that the condition (\ref{13eq:sunbuck}) reported in \cite{sb99} is also conservative.
From Fig.~\ref{Kmax_buck}, the buck converter is susceptible to FSI if $D$ is too small.

In the above analysis, $\wz \ll \ws$ is assumed.
If that is not the case, the loop gain (\ref{eq:highe}) belongs to case $\myci$.
Based on 
Table~\ref{13tab}, the (general) stability condition is 
\begin{equation} 
\mys:= \frac{\va R_s K_c}{T  L  \ws^2}((\ab+(\frac{1}{p}-\frac{1}{z})(\al-\az))) <\phpd
\end{equation}

\subsection{ACMC with PI Compensator: Case $\mycg$ or $\mycb$} 

Let the PI compensator be 
\begin{equation} \label{eq:gc3}
\gc=\frac{K_c (1+s/ \wz)}{s }
\end{equation}
Although the PI compensator is a special case of the type-II compensator by setting $\wpp \to \infty$ in (\ref{eq:gc2}),
here $\wz \ll \ws$ is not assumed as in Sec.~\ref{sec:acmc2} and a separate discussion on the effect of $\wz$ is needed.

Let $\mykk=\va R_s K_c/\vm  L  \ws^2$ (a little different from $K$). Then
\begin{equation}
\myt(s) =  \frac{\va \gc \gi}{\vm}= 
\frac{
\va R_s K_c   (1+\frac{s}{ \wz})}{\vm L s^2 } 
= \frac{\mykk \ws^2 (1+\frac{s}{ \wz}) }{s^2 } 
\label{eq:high22}
\end{equation}

\subsubsection{Based on SSAA: Converter is expected to be stable}
Setting $|\myt(j \wc)|=1$ in (\ref{eq:high22}) leads to
\begin{equation} 
\wc = \frac{\sqrt{2} \ws \mykk z}{\sqrt{\sqrt{\mykk^4+4\mykk^2 z^4} -\mykk^2}}
\end{equation}
For $z^2 \ll  \mykk$, $\wc \approx \ws \mykk /z$ and PM = $\arctan(\wc / \wz)=\arctan(\mykk / z^2) \approx 90^\circ$.
However, FSI may still occur as discussed next.

\subsubsection{Based on HBA: FSI may occur even with PM $\approx 90^\circ$}
From 
(\ref{eq:high22}), $\myt(s)$ belongs to case $\mycg$ 
in
Table~\ref{13tab}, and the stability condition is
\begin{equation} \label{13eq:bc_c7}
\mykk( \frac{\az }{z} +\ab) 
< 1
\end{equation}
Express (\ref{13eq:bc_c7}) in terms of the required ramp slope $\phpd$, as shown in  Table~\ref{23tab2}:
\begin{equation} \label{13eq:bc_c7ma}
\phpd > \mys:=
\frac{\va R_s K_c}{4 L \wz}(4D-2+ (1-2D+2D^2)T \wz )
\end{equation}
For $T \wz \ll 1$ (generally true),
the stability condition (\ref{13eq:bc_c7ma}) becomes 
\begin{equation} \label{eq:acmcpi}
 \phpd  > \mys:= \frac{\va R_s K_c}{L \wz}(D-\frac{1}{2 } )
\end{equation}
agreed with \cite[Eq. 9]{slgz01}.
For $D < 1/2$, the converter is stable even if $\phpd=0$.
For $D > 1/2$, a ramp slope with (\ref{eq:acmcpi}) is required.
A small $\wz$ also makes 
the loop gain (\ref{eq:high22}) belong to $\mycb$ (which has a stability condition like PCMC) 
instead of $\mycg$.
Setting $D=1$ in (\ref{eq:acmcpi}), a (conservative) ramp slope $\phpd=\va R_s K_c/2L \wz$ stabilizes the converter for {\em any} $D$. 

\begin{figure}[t]
    \centering 
\includegraphics[width=0.55\columnwidth]{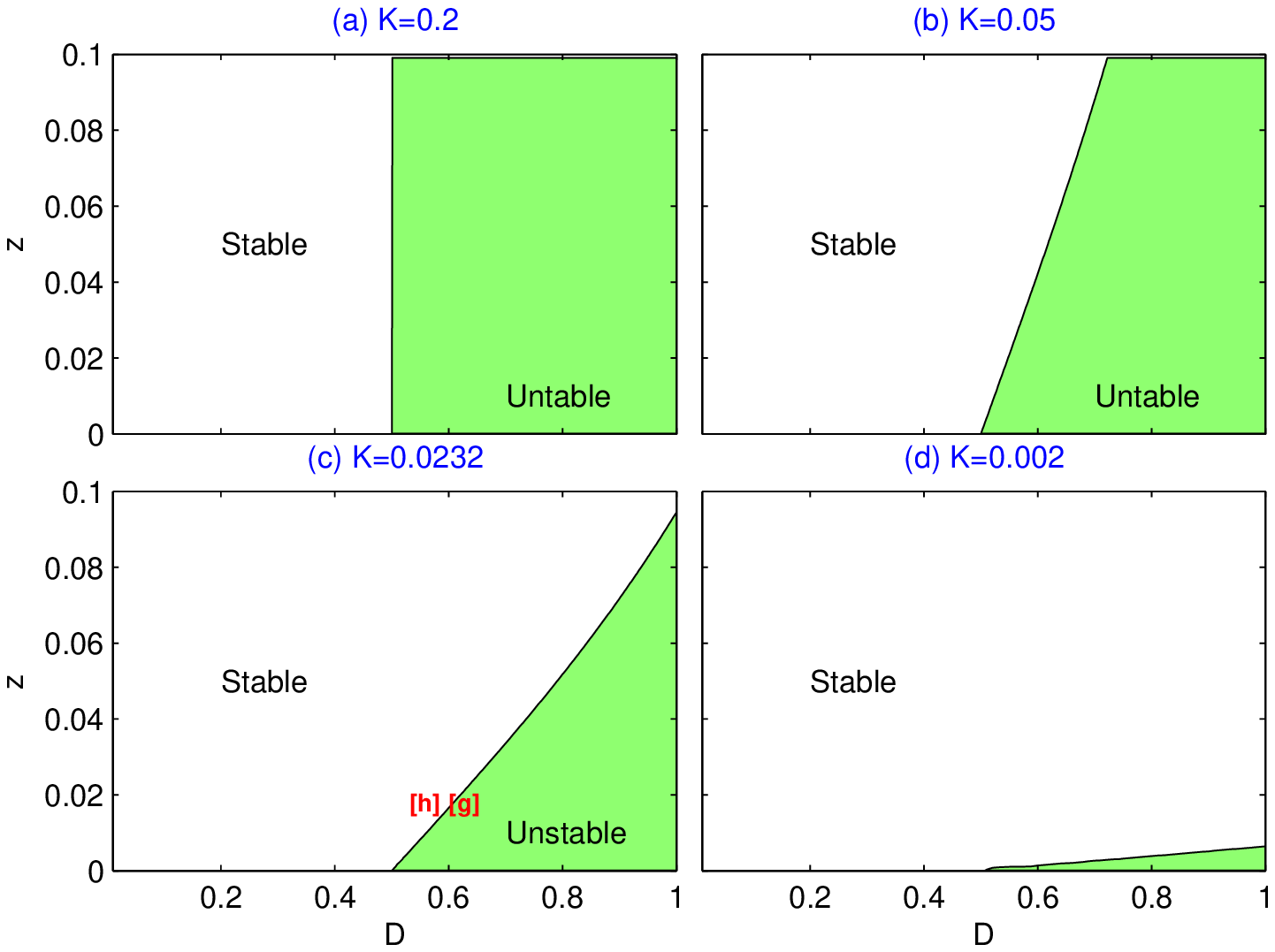}  
\caption{As $K$ decreases or $z$ increases, the stability region enlarges.}
 \label{c7sta}
\end{figure}

For $\mykk=0.2$, 0.05, 0.02, and 0.002, the stable regions are shown in Fig.~\ref{c7sta}.
For $\mykk<0.002$, the whole region in Fig.~\ref{c7sta} is almost stable.
However, FSI still occurs for $D>0.5$ if $z$ is too small.
From Fig.~\ref{c7sta}, the stability is $z$ dependent, even for small $z<0.1$.

\begin{figure}[t]
    \centering 
\includegraphics[width=0.55\columnwidth]{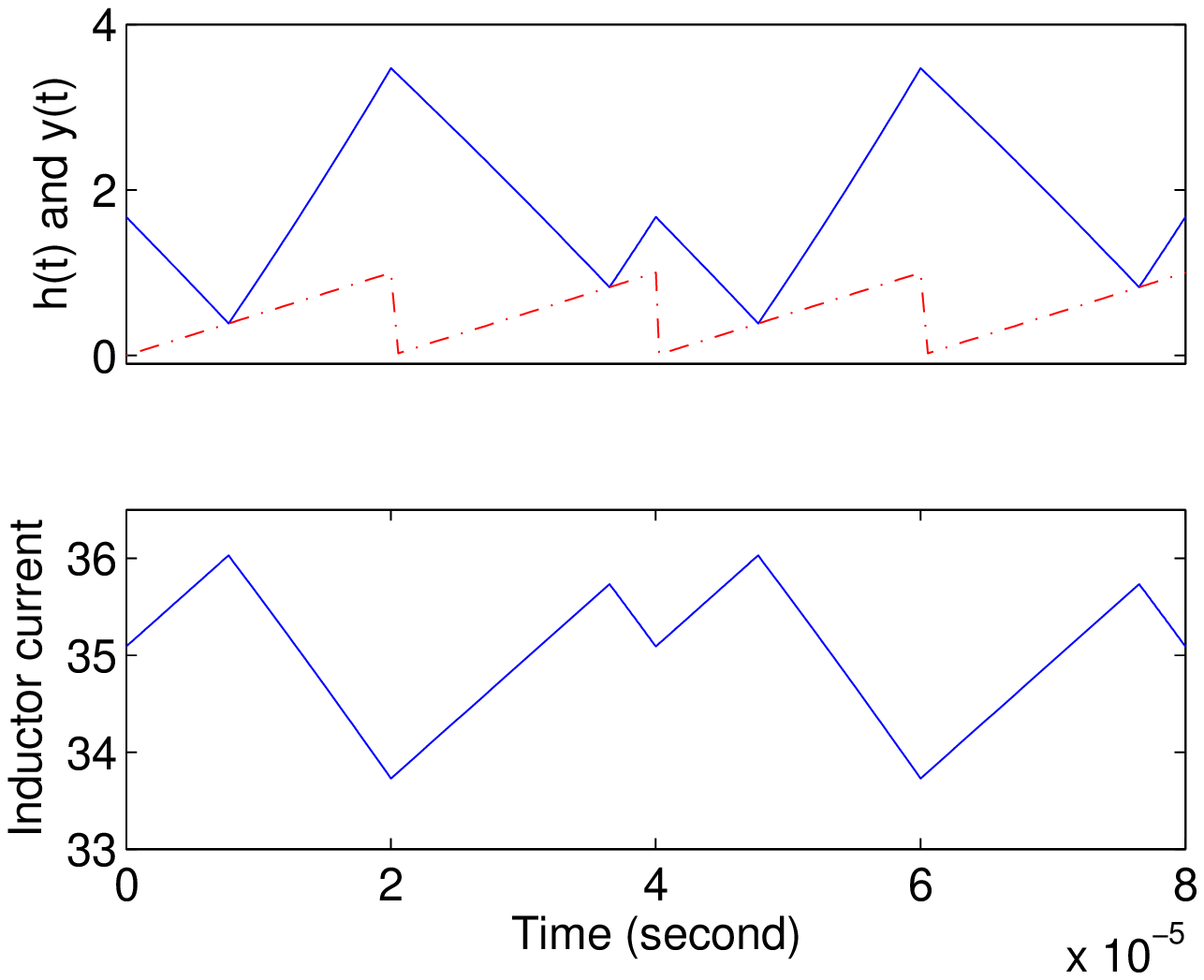}
\caption{The boost converter is unstable, $\vs=5.6$ V.}
 \label{boo_tPID6}
\end{figure}

\begin{figure}[t]
    \centering 
\includegraphics[width=0.55\columnwidth]{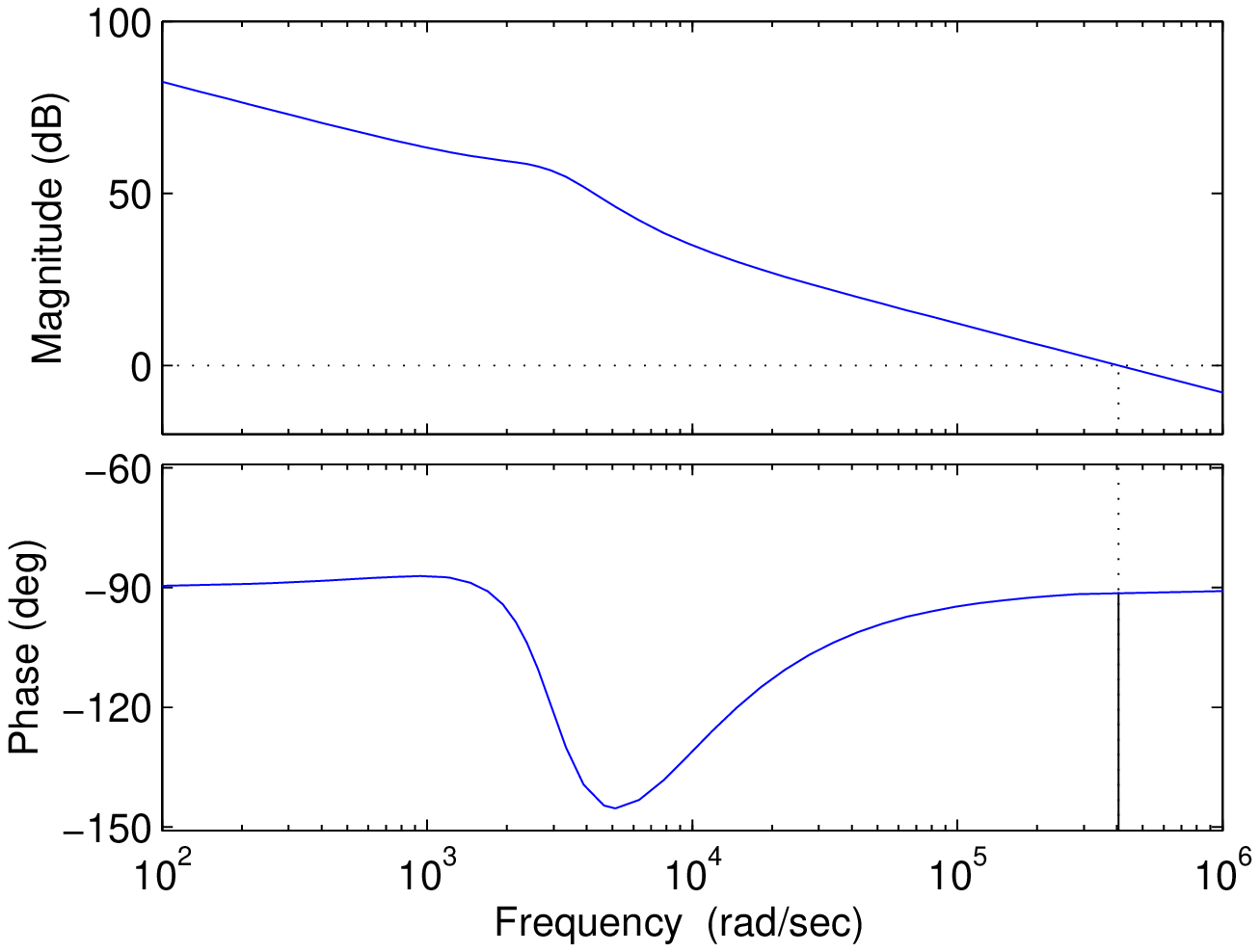}
\caption{The loop gain $\myt(j \w)$ has PM = $89^{\circ}$, $\vs=5.6$ V.}
 \label{boo_PID6}
\end{figure}

\begin{figure}[t]
    \centering 
\includegraphics[width=0.55\columnwidth]{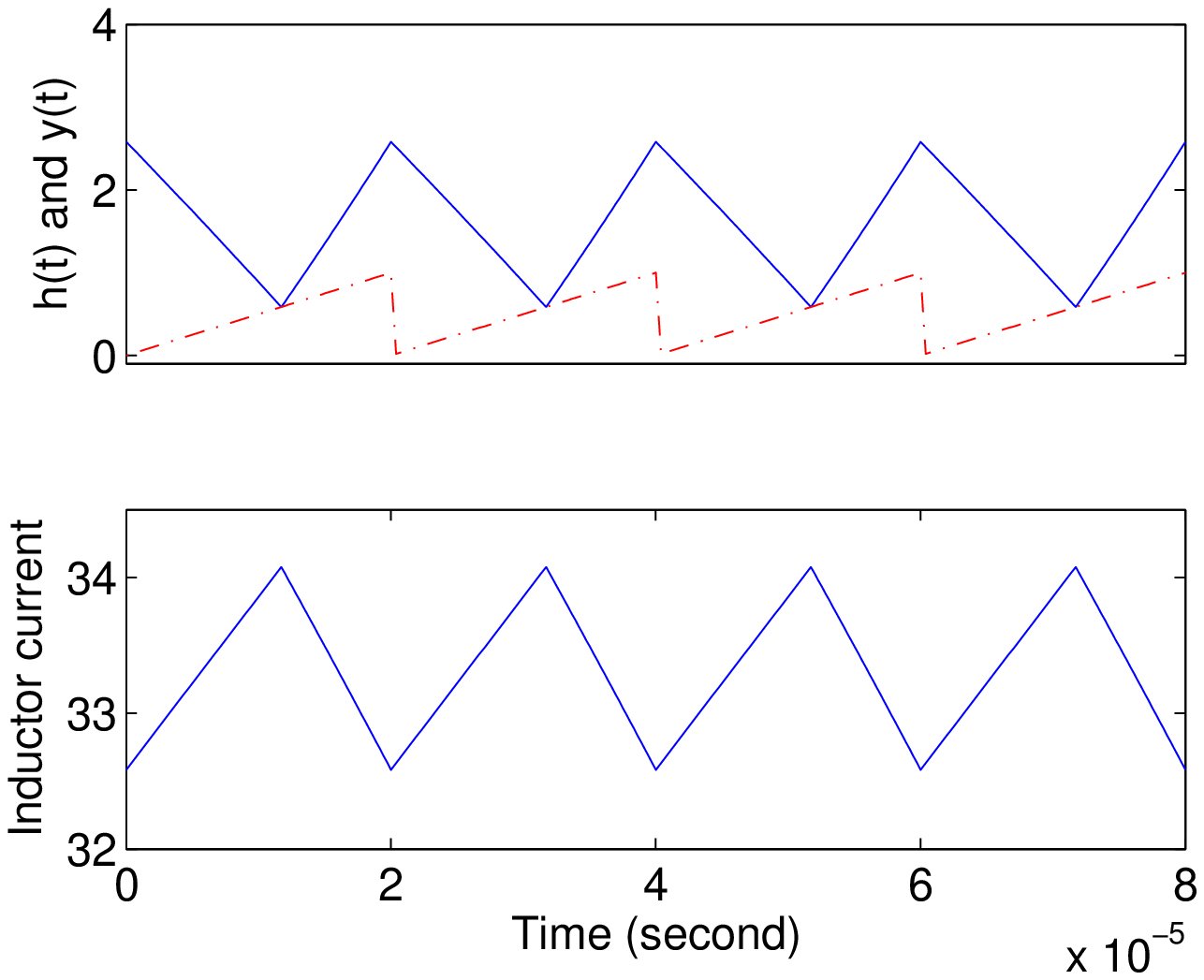}
\caption{The boost converter is stable, $\vs=5.88$ V.}
 \label{boo_tPID58}
\end{figure}

\vspace{2mm} \begin{example} 
({\em FSI with PM = $89^{\circ}$}.)
Continue from Example 1, but with large
$\wpp=3.14 \times 10^9$ rad/s (to make the type-II compensator act like a PI compensator),
$\vs=5.6$ V,
$\vc=0.574$ V, 
and $K_c=460420$.
Here, $z=0.018$, $D=0.6$ and $\mykk=0.0232$.
The converter is unstable (Fig.~\ref{boo_tPID6}) although its average model has PM = $89^{\circ}$ (Fig.~\ref{boo_PID6}).
Independent sampled-data analysis shows 
an unstable pole at -1.02, and three stable poles at 0, 0.88, and 0.91.

Next, let $\vs=5.88$ V and $\vc=0.547$ V. Now, $D=0.58$. The converter is stable (Fig.~\ref{boo_tPID58}).
In Fig.~\ref{c7sta}(c), for $\mykk=0.02$, 
draw a line at $z=0.018$, and the instability indeed occurs around $D=0.6$.
\hspace{3mm} \mbox{  } \hfill  $\Box$  \end{example}

The condition (\ref{13eq:bc_c7}) can be also expressed in terms of $\mykk$:
\begin{equation} \label{13eq:bc_c72}
\mykk 
< \frac{1}{ \az /z+\ab}:=\mykmax(D,z)
\end{equation}
if $\az /z+\ab$  
is positive.

\begin{figure}[t]
    \centering 
\includegraphics[width=0.55\columnwidth]{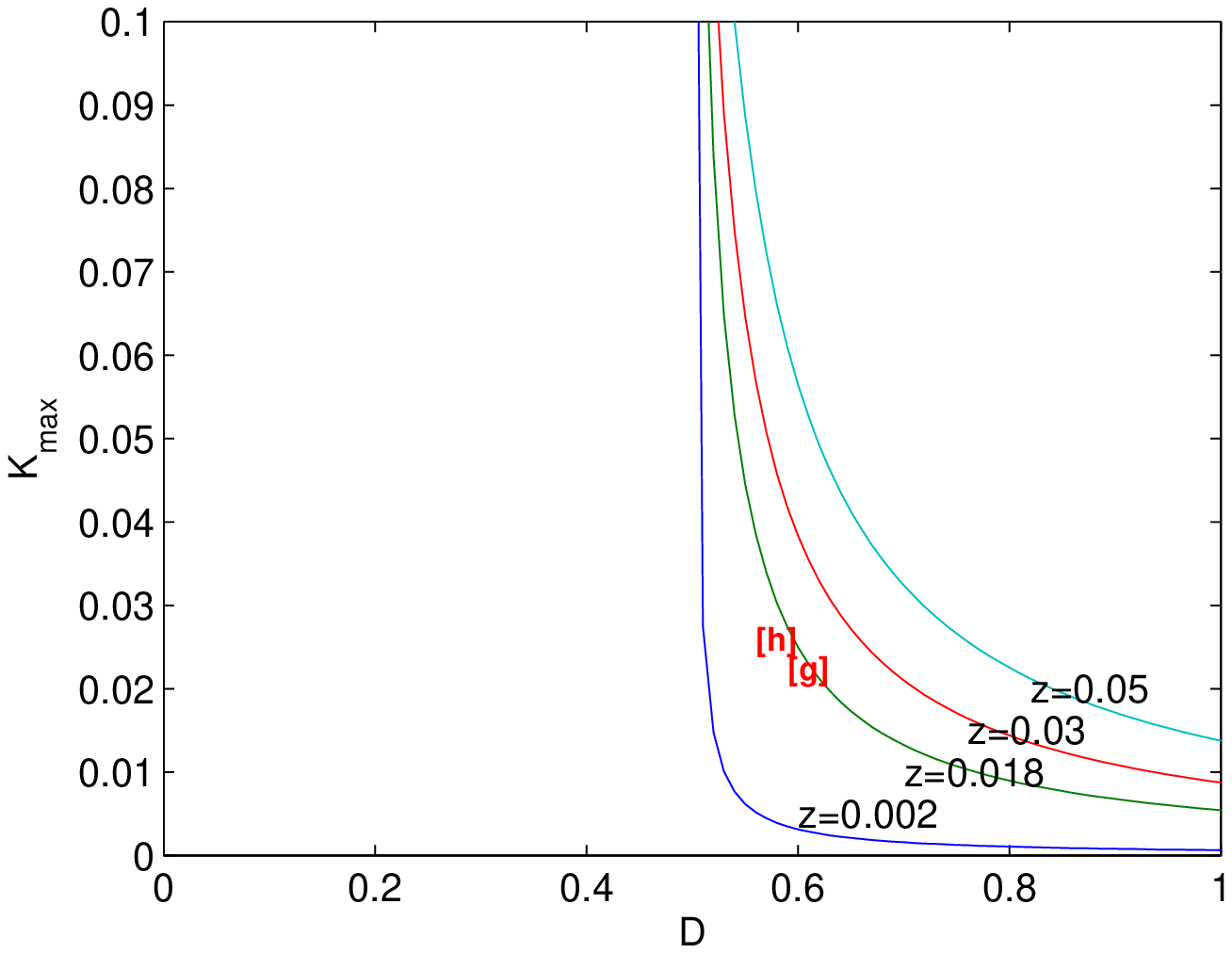} 
\caption{The plots of $\mykmax(D,z)$ for the boost converter.}
 \label{Kmax_booPI}
\end{figure}

For the boost converter, $\va=\vo$, 
and
(\ref{13eq:bc_c72}) becomes
\begin{equation} \label{13eq:bc_c72boo}
\mykk = \frac{ \vo R_s K_c}{\vm  L  \ws^2}
< \mykmax(D,z)
\end{equation}
The plots of $\mykmax(D,z)$ for different values of $z$ are shown in Fig.~\ref{Kmax_booPI}.
For $D< 1/2$, the converter is stable, 
agreed with Fig.~\ref{c7sta}. If $z$ is small, the converter is 
prone to be unstable for $D> 0.5$, agreed with (\ref{eq:acmcpi}). 
As $z$ increases, $\mykmax(D,z)$ increases and the stability 
region enlarges, also agreed with Fig.~\ref{c7sta}.
 The plots of 
$\mykmax(D,z)$ also agree with Example 3. 
Draw a line at $\mykk=0.0232$ in Fig.~\ref{Kmax_booPI}, 
the line intersects  with $\mykmax(D,0.018)$ around $D=0.6$ indicating the onset of FSI at $D=0.6$ as discussed in Example 3.

In Fig.~\ref{Kmax_booPI}, given any value of $z$, $\mykmax(D,z)$ has a minimum at $D=1$.
Then,
a conservative (valid for any $D$) stability condition is
\begin{equation} \label{eq:picon}
\mykk < \mykmax(1,z) =\frac{z}{\pi(1+ \pi z)} < \frac{z}{\pi}
\end{equation}
As discussed above, $\wc \approx \ws \mykk /z$ for $z^2 \ll  \mykk$.
Then, (\ref{eq:picon}) is equivalent to $\wc  < \ws/ \pi $.
This agrees with the tradition wisdom not to set a large $\wc$ to avoid FSI \cite{jnewhb}.
In Examples 1-3, FSI occurs with $\wc  > \ws/ \pi  $.

\section{The Effect of the Voltage Feedback Loop Ripple} \label{13sec:vloop} 
In the above analysis, $\vc$ is assumed constant.
In this section, the effect of  $\vc$ ripple 
generated from the voltage feedback loop is analyzed.
Consider the PCMC buck converter, for example.
Similar analysis can be applied to the ACMC case.

\begin{figure}[!t]  
 \centering 
\includegraphics[128pt,347pt][505pt,484pt]{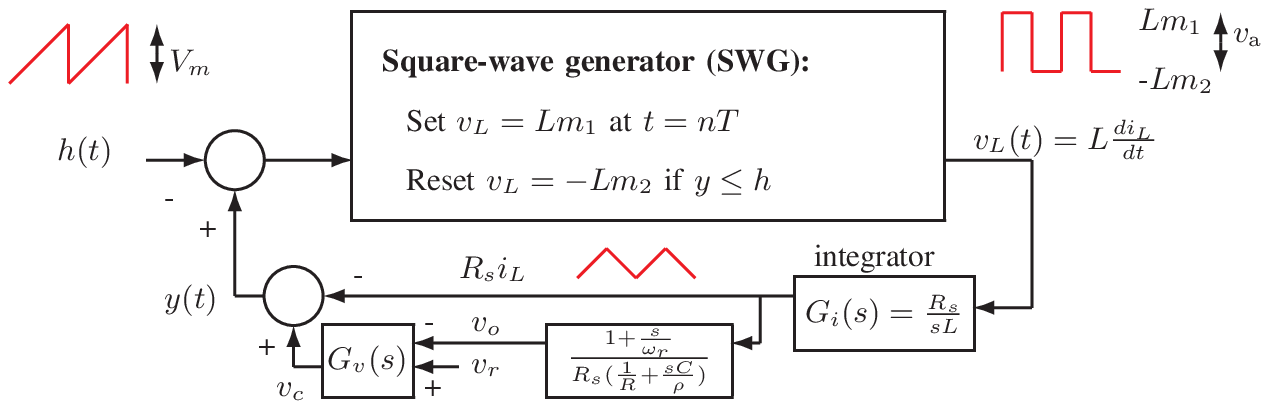}
\caption{A model for PCMC buck converter with closed voltage loop.}
\label{23unified_pcmc} 
\end{figure}

For PCMC, $\gc=1$ and $y=\vc-R_s i_L$ which has two terms, for the voltage and current  loops, respectively.
Let (the ESR zero) $\wesr=1/R_c C$, $r=\wesr / \ws$ and $\rho = R/(R+R_c)$.
Let $G(s):=-y(s)/v_L(s)$.
For the buck converter, 
\begin{equation} 
\frac{\vo(s)}{i_L(s)}=
R \parallel (R_c + \frac{1}{sC})=\frac{1+\frac{s}{\wesr}}{\frac{1}{R}+\frac{sC}{\rho}}
\approx \frac{\rho(1+\frac{s}{\wesr})}{sC} \mbox{   (at high frequency)} 
\end{equation}
Based on Fig.~\ref{23unified} and as shown in Fig.~\ref{23unified_pcmc}, 
the PCMC buck converter can be modeled as 
an SWG plus $G(s)$, where 
\begin{equation} 
G(s) 
=(1+  \frac{\rho  (1+s/\wesr) \gv}{ R_s C s^2} ) \gi
\end{equation}
%
The loop gain is 
\begin{equation} \label{eq:myt_pcmc}
\myt(s) 
=\frac{\va G(s)}{\vm}
=\frac{\vs G(s)}{\vm}
\approx 
 \frac{\vs R_s}{\vm L s}  + \frac{\rho \vs  (1+s/\wesr) \gv}{\vm LC s^2} 
\end{equation}

Three different voltage-loop compensators (with different $\gv$) 
are considered.
The stability conditions  have a {\em universal} form, $\vs R_s (D-0.5) /L < m_a - \mv $, even if different voltage-loop compensators are used, 
and those stability conditions are summarized in Table~\ref{14tab2}.

\begin{table}[!t] 
\caption{Stability conditions for PCMC buck converter with open or closed voltage loop.}
\centering 
\begin{tabular}{ll}
	\hline
 {\bf Universal stability condition:}  & $\frac{\vs R_s (D-0.5) }{L} < m_a - \mv $ \\
\hspace{3mm} {\bf Voltage loop open:} & $\mv=0$ \\
\hspace{3mm} {\bf Voltage loop closed:} & \\
\hspace{6mm} {\bf Proportional gain compensator, $\gv=k_p$:} & $\mv =\frac{\rho \vs k_p }{T LC \ws^2} [\frac{1}{r} \az + \ab]$ \\
\hspace{6mm} {\bf Type-II compensator, $\gv=\frac{K_c (1+s/\wz)}{s (1+s/\wpp)}$:} & $\mv =\frac{K_v}{\ws^2} [\ab +(\frac{1}{p}-\frac{1}{r}) (\al -\az)]$ \\
\hspace{6mm} {\bf PI compensator, $\gv=\frac{K_c (1+s/\wz)}{s}$:} & $\mv =\frac{K_v}{\ws^2} [\frac{1}{r} \az + \ab]$ \\
\multicolumn{2}{l}{Note: $p=\wpp/ \ws$, $z=\wz/ \ws$, $r=\wesr/ \ws=1/R_c C \ws$, $\rho = R/(R+R_c)$, and
$K_v=\rho \vs K_c  / T L C \wz  $}\\
	\hline
\end{tabular}
 \label{14tab2}
\end{table}
\subsection{Proportional Gain Compensator: $\gv=k_p$} 
Let the voltage feedback be $\vc=k_p (\vr-\vo)$. 
From (\ref{eq:myt_pcmc}),
\begin{equation} 
\myt(s) 
= \frac{\vs R_s}{\vm L s}+ \frac{\rho \vs k_p (1+s/\wesr)}{\vm LC s^2}
\end{equation}
 From Table~\ref{13tab}, the stability condition is 
\begin{equation} \label{eq:loop1}
\frac{\vs R_s \az}{\vm L \ws}+ \frac{\rho \vs k_p }{\vm LC \ws^2} (\frac{1}{r} \az + \ab) <1
\end{equation}
or expressed in terms of the ramp slope 
\begin{equation} \label{eq:cmcv}
\frac{\vs R_s (D-0.5) }{L} < m_a - \mv 
\end{equation} 
where, compared with (\ref{13eq:bc_cmc}), the (universal) stability condition (\ref{eq:cmcv}) has an additional term 
\begin{equation} \label{eq:mvkp}
\mv =\frac{\rho \vs k_p }{T LC \ws^2} [\frac{1}{r} \az + \ab]
\end{equation}
due to the effect of the voltage loop ripple.
Note that $\ab > 0$, but $\az < 0$ if $D < 0.5$.
Depending on whether $\mv$ is positive or negative,
the stability region shrinks or enlarges respectively by closing the voltage loop.
It can be proved that for most practical buck converters, $\mv > 0$.

The stability condition (\ref{eq:loop1}) can be also expressed in 
terms of $k_p$, 
\begin{equation} \label{eq:kp}
k_p < \frac{\frac{\ws C}{\rho} (\frac{\vm L \ws}{\vs}- R_s \az) }{\frac{1}{r} \az + \ab}
\end{equation}

\vspace{2mm} \begin{example} 
({\em Accurate prediction of critical gain $k_p^*$}.)
Consider a PCMC buck converter with the voltage loop closed 
from Example 4 of \cite{jcmc}. Simulation and independent 
sampled-data analysis show that FSI occurs at $k_p^* = 237$ (see Fig.~7 of \cite{jcmc}), which can be predicted 
by (\ref{eq:kp}) exactly. In contrary, with $k_p = 237$, the Ridley average 
model  \cite{r91} shows that the converter is stable with an 
infinite gain margin and PM = $36.5^{\circ}$ \cite{jcmc}.
\hfill  $\Box$  \end{example}
\subsection{Type-II Compensator: $\gv=K_c (1+s/\wz)/s(1+s/\wpp)$} 
Let the voltage feedback be $\vc=\gv (\vr-\vo)+\vr$, which has 
an additional offset $\vr$ but it does not affect the loop gain. 
From (\ref{eq:myt_pcmc}),
the loop gain is
\begin{equation} 
\myt(s) 
= \frac{\vs R_s}{\vm L s}+ \frac{\rho \vs K_c (1+s/\wesr)(1+s/\wz)}{\vm LC s^3(1+s/\wpp)}
\end{equation}
Generally $\wz \ll \ws$. Let $K_v 
=\rho \vs K_c  / T L C \wz  $.
 From Table~\ref{13tab}, the stability condition is also (\ref{eq:cmcv}), where
\begin{equation} 
\mv= \frac{K_v}{\ws^2} [\ab +(\frac{1}{p}-\frac{1}{r}) (\al -\az)] 
\end{equation}

\subsection{PI Compensator: $\gv=K_c (1/s+1/\wz)$} 
The PI compensator is a special case of the type-II compensator by setting $\wpp \to \infty$.
Let the control voltage at the output of the voltage-loop compensator be $\vc=\gv (\vr-\vo)+\vr$. 
From (\ref{eq:myt_pcmc}),
the loop gain is
\begin{equation} 
\myt(s) 
= \frac{\vs R_s}{\vm L s}+ \frac{\rho \vs K_c (1+s/\wesr)(1+s/\wz)}{\vm LC s^3}
\end{equation}
Generally $\wz \ll \ws$. 
 From Table~\ref{13tab}, the stability condition is also (\ref{eq:cmcv}), where
\begin{equation} \label{eq:mvpi}
\mv= \frac{K_v}{\ws^2}[\frac{1}{r} \az + \ab] 
\end{equation}
Note that (\ref{eq:mvpi}) is for the PI compensator whereas
(\ref{eq:mvkp}) is for the proportional compensator.
However, they are the same  by setting $k_p=K_c/\wz$.
The proportional compensator, though simple, can be used to predict FSI if a more complicated PI compensator is used.

\section{Conclusion and Contributions
} \label{13sec:conclu}
Based on \cite{jhb12,jnewhb}, a unified CMC model (Fig.~\ref{23unified})  is proposed to predict FSI for different converters under PCMC or ACMC.
Such a unified CMC model exists because any CMC converter is essentially a TWG with a linear feedback.
Closed-form stability conditions are derived (see Table~\ref{23tab2}) and 
verified by time-domain simulations (see Table~\ref{22tab2}).
The obtained results are consistent with (but broader than) the past research such as \cite{slgz01,sb99}.
 The 
instability is found to be associated with large crossover 
frequency. 
A conservative condition to avoid FSI is $\wc < \ws /\pi$.
The proposed model can be applied to 
converters with high-order compensators, such as type-II and PI compensators, for example. 

The questions asked in the Introduction are answered:
\begin{enumerate}
\item FSI occurs in both the buck and the boost converters with the same 
parameters if they have the same $\va$, as shown in Example 2.
\item The {\em unified} model 
can be applied to both PCMC and ACMC. 
\item The {\em same} FSI condition expressed in terms of $\va$, as shown in Table~\ref{23tab2}, also applies to any CMC converter. 
For the buck converter, $\va=\vs=\vo /D$.
For the boost or buck-boost converter, $\va=\vs/(1-D)=\vo$.
For example, given a buck converter with $\vo=\va D$ and a boost converter with $\vo=\va$, if both of the converters have the same power stage parameters, then they have the same stability or instability.
\item Although different parameters have different effects, they can be consolidated into 
a few parameters: $K$, $D$, and $p$. 
A  {\em single} plot of $\kmax(D,p)$ 
can be used to predict FSI.
The stability based 
on traditional average analysis is $D$ independent 
(Fig.~\ref{c5pm}), whereas the actual stability is $D$ dependent 
(Fig.~\ref{c5sta}). 
\end{enumerate}

To the author's knowledge, the following contributions have not been reported:
\begin{enumerate}
\item The {\em unified} CMC model of Fig.~\ref{23unified}, applicable to PCMC or ACMC buck, boost, and buck-boost converters.
\item The {\em unified} stability conditions in Table~\ref{23tab2}.
\item The plots of Figs.~\ref{c5pm}, \ref{c5sta}, \ref{Kmax_boo}-\ref{c7sta}, and \ref{Kmax_booPI},
which are {\em universal} for any CMC converter, and they are not just for specific examples.
\item Using the plot of $\kmax(D,p)$ as a design tool to avoid FSI.
\item The effects of different parameters  on the stability, such as $K$, the compensator pole $\wpp$ and zero $\wz$, as shown in Figs.~\ref{c5sta} and \ref{c7sta}.
\item The conservative stability condition $\wc < \ws/ \pi$ for the CMC converter with a {\em PI} compensator (whereas the same condition for the CMC converter with the {\em type-II} compensator was reported in \cite{jnewhb}).
\item The effect of the voltage loop ripple on FSI (see Table~\ref{14tab2}).
\end{enumerate}

Although this paper focuses on CMC, the proposed 
analysis can be applied to other schemes (such as VMC and constant on-time control).
As reported in \cite{jnewhb}, ACMC with type-II and PI compensators belong respectively to the cases $\myce$ and $\mycg$.
The derived FSI conditions are 
also applicable to these cases. For example, a buck 
converter with
$\mathrm{V}^2$ control belongs to the case $\mycg$  with $\mykk=\vs /\vm LC \ws^2$ and $\wz=1/R_c C$, and the stability condition is exactly (\ref{13eq:bc_c7}).
Also, a buck converter with a type-II, type-III, or phase-lead compensator belongs to the case $\myce$, and the stability condition is exactly (\ref{13eq:bc_c5}).



\begin{thebibliography}{10}

\bibitem{jcmc}
C.-C. Fang.
\newblock Sampled-data poles, zeros, and modeling for current mode control.
\newblock {\em Int. J. of Circuit Theory Appl.}, 41(2):111--127, Feb. 2013.

\bibitem{ctqls08}
Y.~Chen, C.~K.~M. Tse, S.~Qiu, L.~Lindenmuller, and W.~Schwarz.
\newblock Coexisting fast-scale and slow-scale instability in current-mode
  controlled {DC}/{DC} converters: Analysis, simulation and experimental
  results.
\newblock {\em IEEE transactions on circuits and systems I, Regular papers},
  55(10):3335--3348, Nov. 2008.

\bibitem{slgz01}
T.~Suntio, J.~Lempinen, I.~Gadoura, and K.~Zenger.
\newblock Dynamic effects of inductor current ripple in average current mode
  control.
\newblock In {\em Proc. {IEEE} PESC}, pages 1259--1264, 2001.

\bibitem{jhb12}
C.-C. Fang.
\newblock Critical conditions for a class of switched linear systems based on
  harmonic balance: Applications to dc-dc converters.
\newblock {\em Nonlinear Dynamics}, 70(3):1767--1789, Nov. 2012.

\bibitem{jnewhb}
C.-C. Fang.
\newblock Closed-form critical conditions of subharmonic oscillations for buck
  converters.
\newblock {\em {IEEE} Trans. Circuits Syst. I}, 60(7):1967--1974, Jul. 2013.

\bibitem{jnewhb_cot}
C.-C. Fang.
\newblock Closed-form critical conditions of instabilities for constant on-time
  controlled buck converters.
\newblock {\em {IEEE} Trans. Circuits Syst. I}, 59(12):3090--3097, Dec. 2012.

\bibitem{jdivider}
C.-C. Fang and R.~Redl.
\newblock Subharmonic stability limits for the buck converter with ripple-based
  constant on-time control and feedback filter.
\newblock {\em {IEEE} Trans. Power Electron.}, 29(4):2135--2142, Apr. 2014.

\bibitem{fa01}
C.-C. Fang and E.~H. Abed.
\newblock Saddle-node bifurcation and {N}eimark bifurcation in {PWM} {DC}-{DC}
  converters.
\newblock In S.~Banerjee and G.~C. Verghese, editors, {\em Nonlinear Phenomena
  in Power Electronics: Bifurcations, Chaos, Control, and Applications}, pages
  229--240. Wiley, New York, 2001.

\bibitem{em01}
R.~W. Erickson and D.~Maksimovic.
\newblock {\em Fundamentals of Power Electronics}.
\newblock Springer, Berlin, Germany, second edition, 2001.

\bibitem{r91}
R.~B. Ridley.
\newblock A new, continuous-time model for current-mode control.
\newblock {\em {IEEE} Trans. Power Electron.}, 6(2):271--280, 1991.

\bibitem{sb99}
J.~Sun and R.~M. Bass.
\newblock Modeling and practical design issues for average current control.
\newblock In {\em Proc. IEEE APEC}, pages 980--986, 1999.

\bibitem{ylm13}
Y.~Yan, F.C. Lee, and P.~Mattavelli.
\newblock Analysis and design of average current mode control using describing
  function-based equivalent circuit model.
\newblock {\em {IEEE} Trans. Power Electron.}, 28(10):4732--4741, Oct. 2013.

\bibitem{d90}
L.~H. Dixon.
\newblock Average current-mode control of switching power supplies.
\newblock {\em Unitrode Power Supply Design Seminar Handbook}, 1990.

\bibitem{tlr93}
W.~Tang, F.~C. Lee, and R.~B. Ridley.
\newblock Small-signal modeling of average current-mode control.
\newblock {\em {IEEE} Trans. Power Electron.}, 8(2):112--119, Apr. 1993.

\bibitem{slc99}
C.~Sun, B.~Lehman, and R.~Ciprian.
\newblock Dynamic modeling and control in average current mode controlled {PWM}
  {DC}/{DC} converter.
\newblock In {\em Proc. {IEEE} PESC}, pages 1152--1157, 1999.

\bibitem{c00}
P~Cooke.
\newblock Modeling average current mode control.
\newblock In {\em Proc. IEEE APEC}, pages 256--262, 2000.

\bibitem{l08}
R.~Li.
\newblock Modeling average-current-mode-controlled multi-phase buck converters.
\newblock In {\em Proc. IEEE APEC}, pages 3299--3305, Aug. 2008.

\bibitem{lolb09}
R.~Li, T.~O'Brien, J.~Lee, and J.~Beecroft.
\newblock A unified small signal analysis of {DC}-{DC} converters with average
  current mode control.
\newblock In {\em Proc. IEEE ECCE}, pages 647--654, 2009.

\bibitem{lolb11}
R.~Li, T.~O'Brien, J.~Lee, and J.~Beecroft.
\newblock Effects of circuit and operating parameters on the small-signal
  dynamics of average-current-mode-controlled {DC}-{DC} converters.
\newblock In {\em IEEE 8th International Conference on Power Electronics and
  ECCE Asia}, pages 60--67, 2011.

\bibitem{ylm11b}
F.~Yu, F.C. Lee, and P.~Mattavelli.
\newblock A small signal model for average current mode control based on
  describing function approach.
\newblock In {\em Proc. IEEE ECCE}, pages 405--412, 2011.

\bibitem{jasym}
C.-C. Fang.
\newblock Asymmetric critical conditions for peak and valley current programmed
  converters at light loading.
\newblock {\em {IEEE} Transactions on Circuits and Systems-I: Regular Papers},
  2013.
\newblock accepted, available: http://dx.doi.org/10.1109/TCSI.2013.2284178.

\end{thebibliography}
\end{document}